\documentclass[fleqn,usenatbib]{mnras}

\usepackage{mathptmx}

\usepackage[T1]{fontenc}

\DeclareRobustCommand{\VAN}[3]{#2}
\let\VANthebibliography\thebibliography
\def\thebibliography{\DeclareRobustCommand{\VAN}[3]{##3}\VANthebibliography}

\usepackage{graphicx}	
\usepackage{amsmath}	
\usepackage{amssymb}	
\usepackage{gensymb}    

\title[Cool ISM Tracing Evolution of Dusty Local ETGs]{Cool Interstellar Medium as an Evolutionary Tracer in ALMA-Observed Local Dusty Early-Type Galaxies}

\author[D.H.W. Glass et al.]{
David H.W. Glass,$^{1}$\thanks{E-mail: DHWGlass@uclan.ac.uk}
Anne. E. Sansom,$^{1}$
Timothy A. Davis$^{2}$
and Cristina C. Popescu$^{1}$
\\
$^{1}$Jeremiah Horrocks Institute for Mathematics, Physics and Astronomy, University of Central Lancashire, Preston, PR1 2HE, UK\\
$^{2}$Cardiff Hub for Astrophysics Research \& Technology, School of Physics \& Astronomy, Cardiff University, Queens Buildings, Cardiff, CF24 3AA, UK\\
}

\date{Accepted XXX. Received YYY; in original form ZZZ}

\pubyear{2022}

\begin{document}
\label{firstpage}
\pagerange{\pageref{firstpage}--\pageref{lastpage}}
\maketitle

\begin{abstract}

The content and distribution of cool interstellar medium (ISM, $<$30K) can indicate the evolutionary mechanisms that transform late-type to early-type galaxies (ETGs). To investigate this, ALMA observations of $^{12}$CO[2-1] line emission were obtained for five dusty ETGs from a complete sample in low-density environments. Four of the ETGs have massive ($\sim$10$^9$ M$_{\odot}$), extended molecular gas reservoirs with effective radii $\sim$3 – 5 kpc. This work provides a kinematic and structural analysis of these observations, to explore possible evolutionary mechanisms. Axisymmetric or bisymmetric kinematic models were fitted to observations of molecular gas discs, to quantify the dominant structures present and highlight additional structures or asymmetries. Integral Field Unit (IFU) observations of these ETGs were also examined where available. Two of the ETGs, GAMA64646 and 622305, appear to have undergone tidal disturbance leading to molecular gas discs and/or star-forming inner rings. GAMA272990 may have undergone a merger, leading to an elliptical galaxy with an embedded star-forming molecular gas disc. GAMA622429 has probably undergone a minor merger, indicated by asymmetry in molecular gas distribution and disturbance in optical images. The remaining ETG, GAMA177186, was affected by source confusion from an offset source which could be a high-mass, dust- and gas-rich object at high redshift. Overall, it appears that a high proportion of dusty ETGs in low-density environments have massive, extended molecular gas reservoirs, and have undergone some kind of interaction as part of their recent evolution. Secular evolution can then (re-)transform the ETGs from star-forming to passive galaxies.

\end{abstract}




\section{Introduction}


Galaxies have been shown in many studies to exhibit bimodality in properties within certain planes. The earliest example is the classification of elliptical and lenticular early-type galaxies (ETGs) with smooth morphologies, and late-type galaxies (LTGs) with spiral arms or irregular features \citep{Hubble26}. Bimodality is also apparent in optical colour vs absolute magnitude diagrams (CMDs), with LTGs forming a “blue cloud” and ETGs forming a “red sequence” \citep{Baldry04,Bell04,Faber07,Martin07}. Plots of derived star formation rate (SFR) versus stellar mass show star-forming galaxies lying along or near to a star-forming main sequence (SFMS) and other galaxies with reduced star formation rates lying below \citep[e.g.][]{Daddi07,Elbaz07,Noeske07}.

Galaxies also exist in the relatively sparse “green valley” (GV) \citep{Wyder07} between red/blue and star-forming/passive regions, with intermediate properties between ETGs and LTGs. However, some doubt has been cast recently on the existence and nature of the GV, with galaxies in the local Universe exhibiting a continuum of properties \citep{Eales17}.  The GV may be due partly to the effects of colours bunching together in optical surveys. The low density of galaxies in the GV has been attributed to a possible short timescale of evolution ($<$1 Gyr) from LTG to ETG \citep[e.g.][]{Salim14}. However, GV galaxies may also have longer timescales for transition (2 - 4 Gyr) \citep[e.g.][]{Phillipps19}. Also, GV galaxies are more highly represented in sub-mm surveys \citep{Eales18}. The exclusion of GV galaxies with active galactic nuclei (AGN) from galaxy samples may also deepen the apparent GV, by excluding galaxies where AGN happen to be active as part of their duty cycles \citep{Hickox14}.

In spite of the uncertain significance of the GV, it is acknowledged that galaxies within this parameter space are evolving, and many are in transition between LTG and ETG, with passive ETGs as the endpoints of evolution \citep[e.g.][]{Martin07,Salim07,Schiminovich07,Martin18}. A variety of mechanisms are responsible for this transition, some of which lead to relatively rapid change (e.g. minor mergers) with timescales of $\sim$1 Gyr, or slower (secular) change with timescales of several Gyr \citep[e.g.][]{Smethurst17, Smethurst18}. Deposition of molecular gas into an ETG during an interaction \citep[e.g.][]{Davis11} or harassment which stabilises molecular gas against star formation \citep{Kormendy04} can trigger environmentally-driven evolution. Secular evolution is driven more by events within the galaxy, e.g. depletion of molecular gas via the action of bars or feedback from supernovae or AGN, or a reduction in star formation efficiency \citep[e.g.][]{Davis15,Brownson20}. \citet{Man18} provide a summary of mechanisms involved. However, the question of which mechanisms are dominant in forming the ETG population in the local Universe remains open, with various studies indicating different mechanisms as more significant. Depending on the galaxy samples and methods used, studies indicate dominance by merger activity leading to relatively rapid evolution \citep{Baldry04, Kaviraj12} or slower processes such as strangulation \citep{Peng15, Eales17,Phillipps19}. A mixture of fast-acting mergers and slower internal mechanisms could be in operation in parallel, with mergers more significant in less dense environments \citep{Schawinski14,Smethurst17,Deeley20}. Rejuvenation is also possible, with an injection of fresh ISM into an ETG triggering new star formation and re-entry into the GV for a time \citep[e.g.][]{Thomas10,Mancini19}. Different mechanisms can also occur in different regions within the same galaxy. M33 has star formation levels consistent with starbursts around its central region, the SFMS in its main disc and GV in its outer disc, implying that different regions within a galaxy can evolve according to different spatial and temporal scales \citep{Thirlwall20}.

ETGs were once thought to have little or no cool (20 – 30K) interstellar medium (ISM) comprising gas and dust \citep[e.g.][]{Gallagher72,Bregman92}. However, it is now known that some visually-classified ETGs across all environments have significant ISM content, with levels similar to those for LTGs in some cases \citep[e.g.][]{Young11,Smith12,Agius13,Davis19}. The presence of ISM and star formation in ETGs can be used as evidence of evolutionary mechanisms responsible for their ongoing evolution. Both the total content of dust and/or gas and its spatial distribution are useful in this respect, as described below.

ISM content and distribution can be used to identify ETGs which have undergone merger activity during their evolution. This could be major mergers of galaxies with similar mass, which are highly disruptive and can form giant elliptical galaxies  \citep{Xilouris04}, or recent minor mergers involving a substantially smaller ISM-rich galaxy which can leave disc structure intact \citep[e.g.][]{Davis11,Kaviraj12}.  A lack of trend in dust mass with stellar mass in local ETGs within the ATLAS-3D survey \citep{Cappellari11} indicates external addition of dust, rather than via internal processes which would form a trend \citep{Smith12}. Kinematic misalignment between cool molecular gas and stellar mass within galaxies can be evidence of merger activity \citep{Davis11}, although further analysis is needed to identify misalignment associated with other effects such as relatively slow accumulation of gas within a disc over long periods following a merger \citep{vandevoort15}. 

Conversely, trends in ISM content with other properties, symmetrical distribution or good kinematic alignment between ISM and stellar mass can be evidence for evolution either via slower internal (secular) processes which do not disturb alignments significantly, or via merger activity further back in time after which ISM has settled into alignment. A rapid addition of fresh ISM to a galaxy could be expected to settle into alignment with stellar mass within a few dynamical timescales e.g. $\sim$100 Myr for the inner parts of ETGs \citep{Tohline82}, although asymmetries have been found observationally to persist beyond this timescale \citep{vandevoort18}.

To investigate further how the distribution of ISM in ETGs can be used as an indicator of evolutionary history, five dusty ETGs were observed at frequencies around the $^{12}$CO[2-1] emission line (230.6 GHz rest frequency) using the Atacama Large Millimeter Array (ALMA) in 2016 \citep[][referred to as S19]{Sansom19}. The target ETGs were chosen from a complete ETG sample \citep{Agius13} derived from the Galaxy and Mass Assembly (GAMA) survey \citep{Driver09}, with dust emission at sub-mm wavelengths detected by the Herschel Space Observatory \citep{Pilbratt10} within the Herschel-ATLAS survey \citep{Eales10}. The dust masses of the target ETGs are amongst the largest in the sample, and are challenging to explain without addition of ISM from external sources \citep{Rowlands12,Smith12}.  Earler studies of ETGs \citep{Young11,Davis13} suggested that ISM was concentrated in disc-like structures, containing molecular hydrogen masses in the range 10$^{7.3 - 9.3}$ M$_{\odot}$ with radii around half the effective optical radius ($\sim$1 - 2 kpc). However, three of the ETGs (GAMA64646, 272990 and 622429) observed with ALMA were found to contain massive molecular gas reservoirs ($\sim$few x 10$^9$ M$_\odot$) with a greater spatial extent than assumed (S19). A weak centrally-concentrated continuum emission source was detected in one ETG (GAMA177186), with another unresolved source of continuum and line emission offset from the centre.  A source of line emission offset from the centre was detected in the remaining ETG, GAMA622305.

S19 describes the sample selection, the observational methods and the findings of the observations. In this paper, the observational findings are analysed to investigate the structure and kinematics of the molecular gas. The results are used to infer likely evolutionary mechanisms at work to form and maintain the detected massive molecular gas reservoirs. This is achieved by fitting axisymmetric or bisymmetric kinematic models to the ALMA observations using the Python implementation of the Kinematic Molecular Simulation (\textsc{KinMSpy}) code\footnote{https://github.com/TimothyADavis/KinMSpy}  \citep{Davis13}, available via the \textsc{KinMS\_fitter}\footnote{https://wisdom-project.org/codes/KinMS\_fitter/} wrapper. The aim is to highlight any asymmetry present, explore the alignment between ISM and stellar mass, model the molecular gas distribution and kinematics, and where possible investigate the potential for star formation within the molecular gas reservoirs. 

The paper is structured as follows. Section 2 describes the galaxies observed and the methods used to analyse the observational data. Section 3 describes the results of kinematic modelling of the observations for the three ETGs with molecular gas reservoirs. Section 3 also compares the ALMA observations with data from Sydney-AAO Multi-object Integral-field unit (SAMI) Galaxy survey DR3 \citep{Croom12,Croom21}, to investigate kinematic alignments with stars and ionised gas and the spatial relation between star formation and molecular gas. Star formation maps of stellar and ionised gas line-of-sight velocities from SAMI are also used for comparison, along with their maps of star formation rate per spaxel. Analyses of other ALMA detections for the remaining two ETGs are also included in Section 3. Section 4 provides a discussion, and conclusions are summarised in Section 5. We assume flat $\Lambda$CDM cosmology, with H$_0$ = 70 km s$^{-1}$ Mpc$^{-1}$, $\Omega_\text{M}$ = 0.3 and $\Omega_\Lambda$ = 0.7. Throughout this paper, continuum milimetre-wavelength emission detected by ALMA is assumed to be from the Rayleigh-Jeans spectrum of cool dust unless stated otherwise.


\section{Data and Methods}
\label{sec:Data_Methods}


\subsection{Observational Data}
\label{sec:ObsData}


Selections for dusty ETGs for ALMA observation are described in S19. The ALMA-observed ETGs (GAMA64646, 177186, 272990, 622305, 622429) are from a complete sample within the GAMA equatorial regions \citep{Agius13}. All are in low-density environments, with surface densities to the fifth nearest neighbour in the range 0.06 - 0.4 Mpc$^{-2}$ from the GAMA DR3 \citep{Baldry18} catalogue EnvironmentMeasuresv05. They were selected as being amongst the dustiest within the ETG sample, with estimated Herschel-detected dust masses in the range 2 – 8 $\times$ 10$^7$ M$_\odot$ (S19, see their figure 2). They were therefore expected to contain significant masses of cool molecular hydrogen, based on previously determined gas-to-dust mass ratios for ETGs \citep[e.g.][]{Smith12}. The ETGs were also selected to have ellipticity $>$ 0.2 to ensure that line-of-sight velocity profiles can be recovered throughout the ETGs, and effective optical radii of 4” $<$ Re $<$ 10” to maximise the likelihood of observing ISM distribution with ALMA with a single pointing. An emission line width of 150 km s$^{-1}$ was assumed for observation planning to estimate emission line strengths, along with an assumed ISM spatial extent of half the optical effective radius \citep{Davis13}.

Selection of the observed ETGs from their original parent sample was partly on the basis of smooth morphology and the absence of strong active galactic nucleus (AGN) activity \citep{Agius13}. Table \ref{tab:Basic_Props} shows key properties of the ETGs from GAMA DR3 catalogues. Since these ALMA observations, deeper and sharper optical images have become available from the Kilo-Degree Survey \citep[KiDS,][]{deJong13}. Figure 1 shows r-band, log-projection, optical images from KiDS, for the five ALMA-observed ETGs, and the ALMA-detected molecular gas and dust (S19). Two of the ETGs (GAMA64646, 622305) have faint spiral structure, which is apparent in the deeper KiDS images but not in Sloan Digital Sky Survey \citep[SDSS,][]{York00} images used originally for morphological classification. In addition, optical line emission diagnostics \citep{cidfernandes10,cidfernandes11} derived from more recent GAMA DR3 catalogues \citep{Baldry18} show that GAMA622429 is associated with strong AGN activity (Figure \ref{fig:WHAN}). Nonetheless, these galaxies are of interest in this study to examine evolutionary mechanisms at work to form ETGs. GAMA64646 and 272990 are shown in Figure \ref{fig:WHAN} as having weak AGN activity, but this has not affected the detection of widely-distributed massive molecular gas reservoirs. None of the five targets are in the star-forming region of Figure \ref{fig:WHAN}.

\begin{figure*}
   \includegraphics[width=\textwidth]{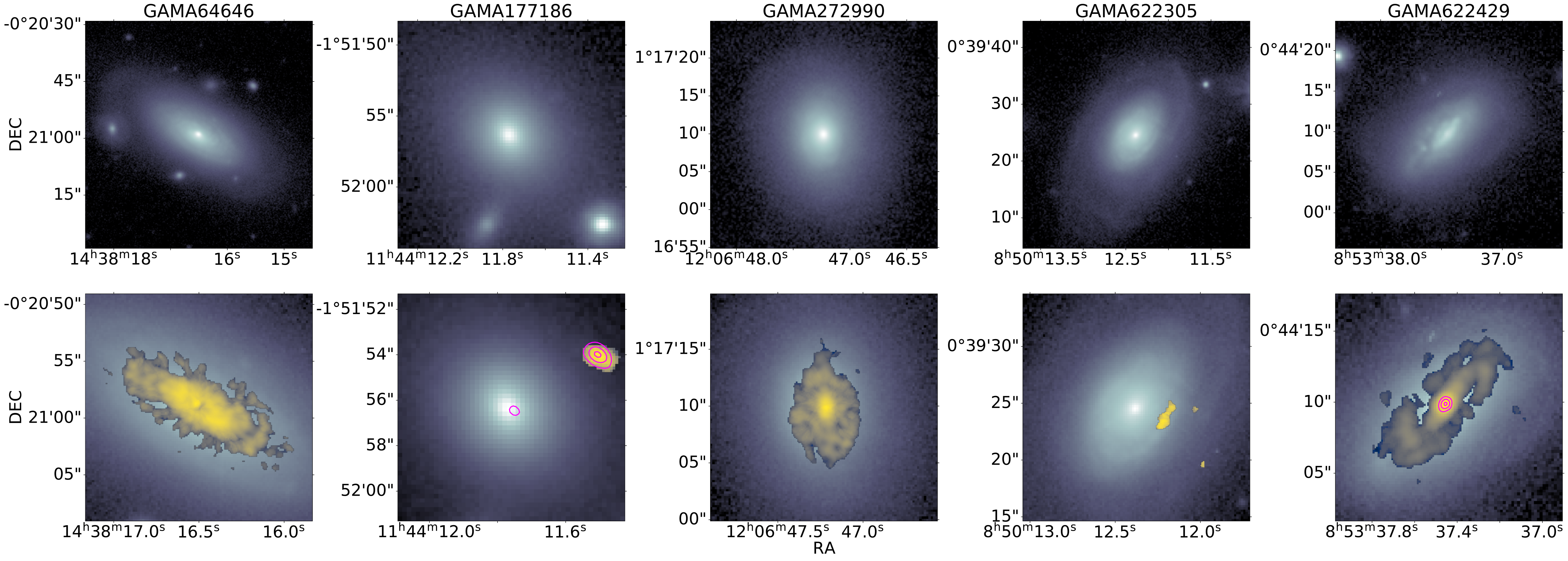}
   \caption{\textit{Upper}: log-normalised KiDS r-band images of the ALMA-observed ETGs. \textit{Lower}: zoomed in, ALMA-detected dust emission (magenta contours) and log-normalised $^{12}$CO[2-1] emission (yellows) overlain.}
    \label{fig:r_gas_dust_overlay}
\end{figure*}

\begin{figure}
   \includegraphics[width=\columnwidth]{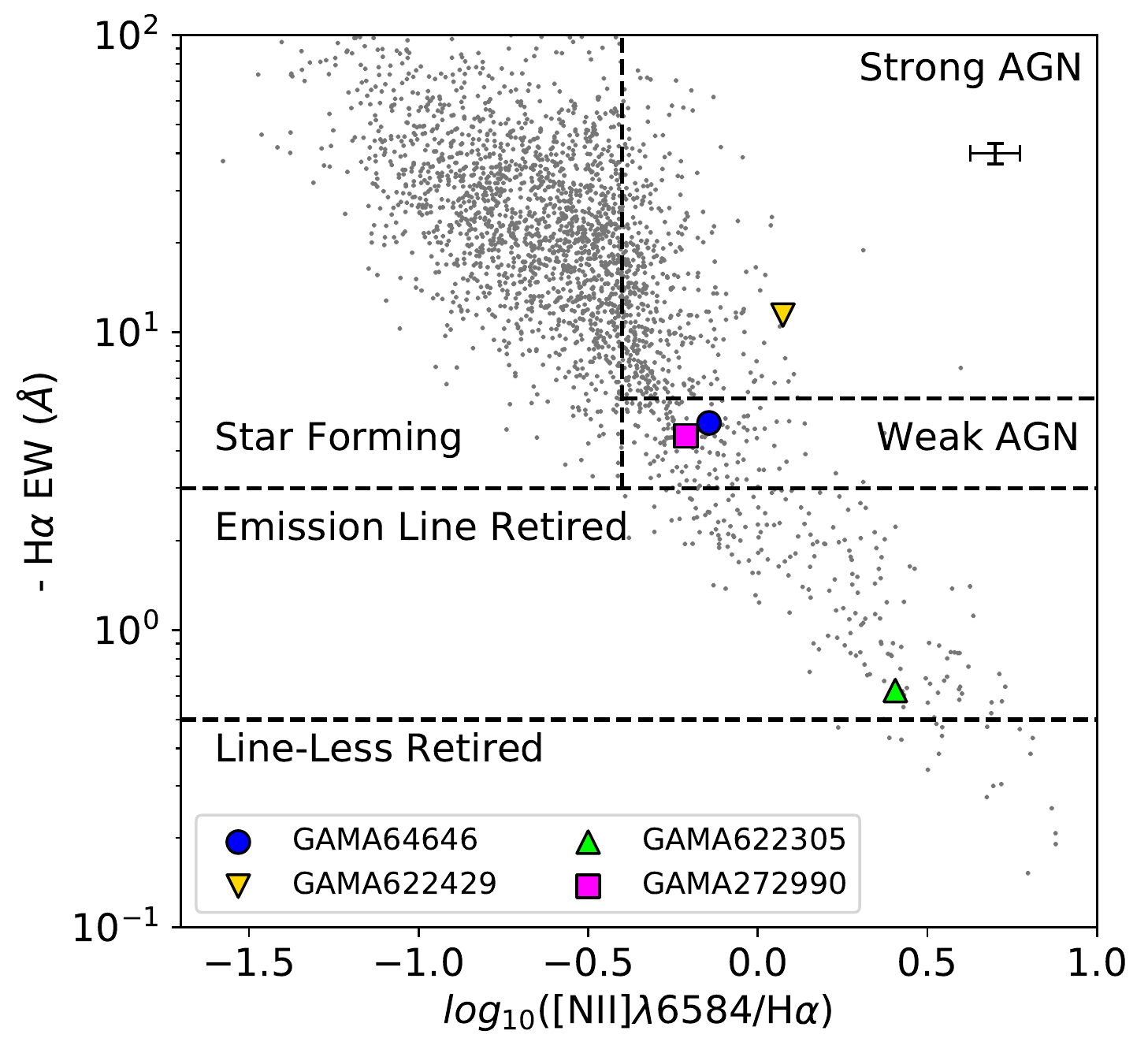}
   \caption{Diagnostic (WHAN) plot for AGN activity, star formation and galaxy retirement using H$_{\alpha}$ equivalent width vs [N2]$\lambda$6584/H$_{\alpha}$ \citep{cidfernandes10,cidfernandes11}, showing the characteristics of the ALMA-observed ETGs. GAMA177186 is not shown because its spectrum does not include suitable emission lines (i.e. it is a line-less retired galaxy). Grey points show a complete galaxy sample from the GAMA equatorial regions (redshift z$\le$0.06, Glass et al., in prep.). Demarcations and terminogy for regions on the diagram are from \citet{Herpich16}.}
    \label{fig:WHAN}
\end{figure}

Figure \ref{fig:r_gas_dust_overlay} also shows that GAMA64646 has a faint tidal tail to the left, which is not associated with galaxy GAMA64647 (redshift $\sim$0.12) at the apparent end of the tail. GAMA272990, although classified as an elliptical using SDSS images, appears to have faint disc-like structure, slightly asymmetric, with faint ring features within it. GAMA622305 appears to be elongated towards the lower left of the image, and GAMA622429 has signs of disturbance within lower portion of the disc. 

\begingroup
\renewcommand{\arraystretch}{1.3} 
\begin{table*}
	\centering
	\caption{Properties of the ALMA-observed ETGs. Optical effective radii from GAMA DR3 were obtained from fitting a single S\'ersic surface brightness profile to optical images. Dust mass estimates from GAMA DR3 were estimated using Magphys \citep{daCunha08}. Detected molecular gas masses are from S19. The molecular gas mass for GAMA177186 is for the offset object, assuming $^{12}$CO[2-1] emission at the redshift of GAMA177186.}
	\label{tab:Basic_Props}
	\begin{tabular}{lcccccc}
		\hline
            \vspace{-0.1cm}
		ETG & Morphology & Redshift                & log$_{10}$(Stellar      & log$_{10}$(Molecular       & log$_{10}$(Dust            & Optical (r-band) \\
                    &                  & (Heliocentric)         &  Mass (M$_{\odot}$))  & Gas Mass (M$_{\odot}$))      &  Mass (M$_{\odot}$))     & Effective Radius (arcsec) \\
            \hline
		GAMA64646   & S0-Sa  & 0.0369  & 10.89$^{+0.00}_{-0.08}$  &  9.51$\pm$0.03                 &  7.54$\pm$0.05 & 9.92$\pm$0.06  \\
		GAMA177186 & E         & 0.0476  & 10.30$^{+0.08}_{-0.15}$  &  8.54$\pm$0.06                 & 7.23$^{+0.08}_{-0.13}$ & 3.19$\pm$0.10  \\
             GAMA272990 & E        & 0.0411  & 10.53$^{+0.08}_{-0.05}$   &  9.38$\pm$0.03                 & 7.38$^{+0.07}_{-0.04}$ & 3.80$\pm$0.03  \\
		GAMA622305 & S0-Sa  & 0.0426  & 10.62$^{+0.01}_{-0.04}$   &  7.81$^{+0.08}_{-0.10}$  & 7.63$^{+0.06}_{-0.08}$ & 4.66$\pm$0.02  \\
             GAMA622429 & S0-Sa  & 0.0409  & 10.51$^{+0.13}_{-0.04}$   &  9.85$\pm$0.03                 & 7.63$^{+0.01}_{-0.00}$ & 4.31$\pm$0.03  \\
		\hline
	\end{tabular}
\end{table*}
\endgroup

The ALMA-detected massive molecular gas reservoirs within three of the ETGs (GAMA64646, 272990, 622429) have greater detected spatial extents (11 – 18”) than predicted at the observation planning stage. A less massive region of molecular gas was detected in GAMA622305, located off-centre by $\sim$4 arcsec but coincident with the galaxy disc along the line of sight. Emission line widths (400 – 500 km s$^{-1}$) are wider than than predicted but not unexpected \citep[e.g.][]{vandevoort18}. Line emission was detected in the observation for GAMA177186, but in a compact region $\sim$4 arcsec from the centre.

Dust detection was challenging with all these observations, because of the maximum recoverable angular scale (MRS) of $\sim$10 arcsec achieved by the array configuration used. Any dust distribution with greater angular scales than this would not be detected. The surface brightness of the total dust emission is reduced if the dust is more widely distributed than assumed during observation planning, also making detection by ALMA more difficult. Centrally-concentrated continuum emission (presumably dust) was detected for GAMA177186 and GAMA622429, and a compact region of contiuum emission was detected in the same location as line emission in the observation of GAMA177186 (S19).

Observations and data from SAMI are also used in this work. Kinematic maps derived from optical spectra for stars and ionised gas are available for all ALMA-observed ETGs except GAMA177186, which allow the alignment of molecular gas and these components to be assessed. Maps of derived SFR are also available, for comparison with molecular gas location and predicted star formation potential from molecular gas properties (Sections \ref{sec:Toomre} and \ref{sec:Kinematic_Alignments}). These SFR maps are acknowledged as being clean but not complete, such that confirmed regions of star formation are highlighted but others may have been missed. This is because methods to separate star formation effects in optical spectra from other excitation mechanisms are not robust \citep{Medling18}.


\subsection{Kinematic Modelling}
\label{sec:Kinematic_Modelling}


Idealised axisymmetric kinematic models were constructed for the three detected massive molecular gas reservoirs using \textsc{KinMSpy}, to highlight deviations from symmetrical distributions within the data and to provide information on rotational velocity and velocity dispersion. This code creates simulated data cubes representing those produced from ALMA observations, by distributing line emission over numerous point sources ("cloudlets") with equal emission. Initial modelling used point sources distributed according to axisymmetric S\'ersic surface brightness profiles for disc-like structures (Equation \ref{eq:SBProf}, \citet{Sersic63}). 

\begin{equation}
    \ln\left(S\right)=\ln\left(S_0\right)-\left(\frac{R}{r_0}\right)^n
	\label{eq:SBProf}
\end{equation}

where S is surface brightness at radius R, S$_0$ is surface brightness at the centre, r$_0$ is a radial scaling factor and n is the S\'ersic index. 

The point-source distributions were then oriented to a given position angle and ellipticity, and subjected to an axisymmetric circular velocity profile to generate a synthetic 3-dimensional (position + frequency) emission data cube. Equation \ref{eq:velprof} shows the empirical arctangent-based model for radial velocity used, which gives a rapid rise in rotation velocity from the galactic centre followed by a relatively constant velocity \citep[e.g.][]{vandevoort18}.  

\begin{equation}
    v_c=\left( \frac{2v_{\rm flat}}{\pi} \right)\arctan\left( \frac{R}{R_0} \right)
	\label{eq:velprof}
\end{equation}

where $v_c$ is the circular velocity (km s$^{-1}$) at radius $R$ (arcsec), $v_{\rm flat}$ is the far-field circular velocity  (km s$^{-1}$) and $R_0$ is a radial scaling factor (arcsec).  In some cases the fitted far-field circular velocity may be greater than the maximum observed velocities e.g. derived from spectra, because the velocity profile in Equation \ref{eq:velprof} continues to increase weakly beyond the extent of the galaxy. Velocity dispersion was also applied to the model, in this case as a spatially constant value everywhere in the molecular gas disc. Finally the models were convolved with a synthesised beam derived from the ALMA observation, to allow comparison of model and data.

To avoid the fitting of models to observational data away from the ETGs, where primary beam correction during ALMA data reduction may amplify noise towards the edge, a unique elliptical mask per ETG capturing the $^{12}$CO[2-1] emission was applied to each velocity frame for data and model before calculating $\chi^2$ or log likelihood. A curve-of-growth approach was used to optimise the spatial dimensions of elliptical masks. Initial elliptical masks were created based on the dimensions and position angles of molecular gas in zeroth-order moment maps, and a multiple of the dimensions was found which led to a maximum total flux while minimising the mask area. Table \ref{tab:Ellipse_Masks} shows the resultant dimensions and position angles of the masks, along with the flux values determined. Uncertainties in the flux values include 6\% calibration uncertainty for the ALMA flux calibrators added in quadrature (see https://almascience.eso.org/sc/). The total fluxes are in agreement with the fluxes derived in S19 within errors, so the molecular gas masses in Table \ref{tab:Basic_Props} are unchanged.

\begin{table}
	\centering
	\caption{Optimal ellipse mask sizes and ALMA-observed $^{12}$CO[2-1] emission fluxes for GAMA64646, 272990 and 622429.}
	\label{tab:Ellipse_Masks}
	\begin{tabular}{lcccc}
		\hline
		Galaxy & Major & Minor & Position & $^{12}$CO[2-1]  \\
                       & Diameter & Diameter & Angle & Flux \\
                       & (arcsec) & (arcsec) & ($\degree$) & (Jy km s$^{-1}$) \\
            \hline
            GAMA64646   & 19.2 & 12.0 & 62 & 35.0 $\pm$ 3.5 \\
            GAMA272990 &  17.3 & 9.4 & 3 & 21.0 $\pm$ 1.4 \\
            GAMA622429 &  13.0 & 10.3 & 135 & 61.9 $\pm$ 3.9 \\
	\end{tabular}
\end{table}

The parameters for a model data cube can be adjusted to fit an ALMA observation. Initial fits were achieved using \textsc{scipy.optimize.minimize} to find a minimal $\chi^2$. The Nelder-Mead algorithm \citep{Nelder65} was used for fitting, which is a direct search method for the minimum result within a multidimensional parameter space which avoids the need to compute derivatives. A minimal rectangular volume for fitting was selected around the emission region, to contain the optimal elliptical masks determined by curve-of-growth (Table \ref{tab:Ellipse_Masks}). A representative RMS noise per pixel (in Jy beam$^{-1}$) was determined from velocity frames outside of the emission region but within the optimal elliptical mask. The number of point sources for axisymmetric disc components was set to a default value of 100,000 within \textsc{KinMS}, which was found to give similar results to simulations with greater numbers of points while retaining smoothness in zeroth order moment maps. The centroids for the spatial and velocity axes were also fitted, to ensure that uncertainties in these parameters are correctly reflected in uncertainties for others. These were found to have uncertainties less than half the size of the pixels or velocity bins.

The Python implementation of \textsc{GAStimator}\footnote{https://github.com/TimothyADavis/GAStimator}, designed for use with \textsc{KinMSpy}, was then used for Markov-Chain Monte Carlo (MCMC)-based fitting, to refine the fit obtained using \textsc{scipy.optimize.minimize} and to determine posterior distributions for the model parameters. Flat priors were used, with wide but physically realistic upper and lower limits. Initial estimates for parameters were based on fitted parameters from scipy.optimize.minimize.  Convergence of models fitted with \textsc{GAStimator} was assured by inspection of trace plots (fitted value vs. step number) for each parameter, to ensure that the parameter space was frequently sampled over the prior range. If serial correlation was detected, with meandering trace plots and only a few major peaks and troughs, fitting was repeated with an increased number of steps. This was an issue for parameter pairs with degeneracy. Plots of autocorrelation parameter calculated using tools within \textsc{EMCEE} \citep{FM13, FM18} vs number of steps were also used, to ensure that the number of model steps used was sufficient to achieve small autocorrelation parameters ($\sim$0) for all fitted parameters and to select a suitable burn-in interval for later analyses.

When fitting models to data with many points (in this case, up to $\sim$600,000), the variance in $\chi^2$ becomes large if the noise within the data is used for fitting \citep{vandenbosch09}, resulting in extremely narrow posterior distributions. The approach taken to avoid this was to scale the noise to achieve uncertainties that take the variation in $\chi^2$ into account. A 2-stage approach was used, which first scales down the RMS noise within the data to achieve a reduced $\chi^2$ of $\sim$1 for the model fit (using the result from scipy.optimize.minimize) and then multiplies it by an additional factor of $\left(2N\right)^{1/4}$, where N is the number of data points being fitted \citep{Mitzkus17, Smith19}. The use of the number of fitted pixels for N  ensures that the factor is conservatively large, because the pixels oversample the spatial plane in the data compared to a beam area of $\sim$36 pixels. Estimates of parameter uncertainties from model fitting are then conservatively large. It is also possible to apply a covariance matrix within the calculation of log likelihood, which accounts for the correlated uncertainty between pixels A detailed discussion of this approach is provided by \citet{Tsukui22}, and \citet{Smith19} apply this to their model fitting as well as the inflation of RMS noise described above.  However the number of elements within this covariance matrix is N$^4$, and the computer memory needed to invert it becomes rapidly excessive with increasing number of pixels\footnote{https://web.ipac.caltech.edu/staff/fmasci/home/astro\_refs/PixelNoiseCorrelation.pdf}. In practice the contribution to errors from variance in $\chi^2$ dominates over correlated pixel-to-pixel variation \citep{Davis20}, so the covariance matrix is not used.


\section{Results}
\label{sec:Results}



\subsection{Kinematic Modelling}
\label{sec:Results_KinMS}

\begin{figure*}
   \includegraphics[width=\textwidth]{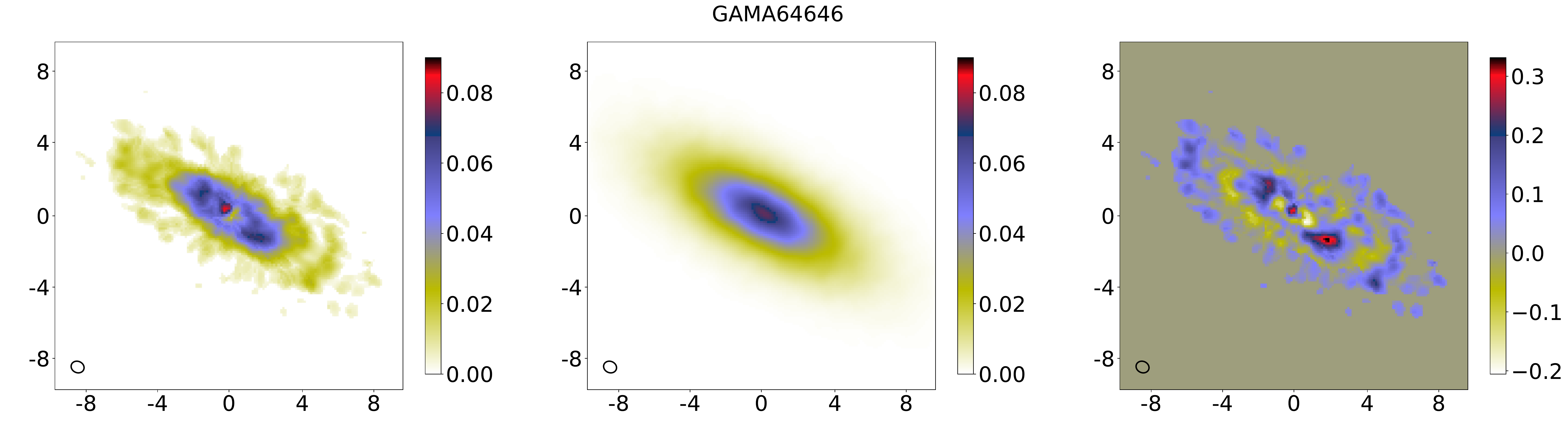}
   \includegraphics[width=\textwidth]{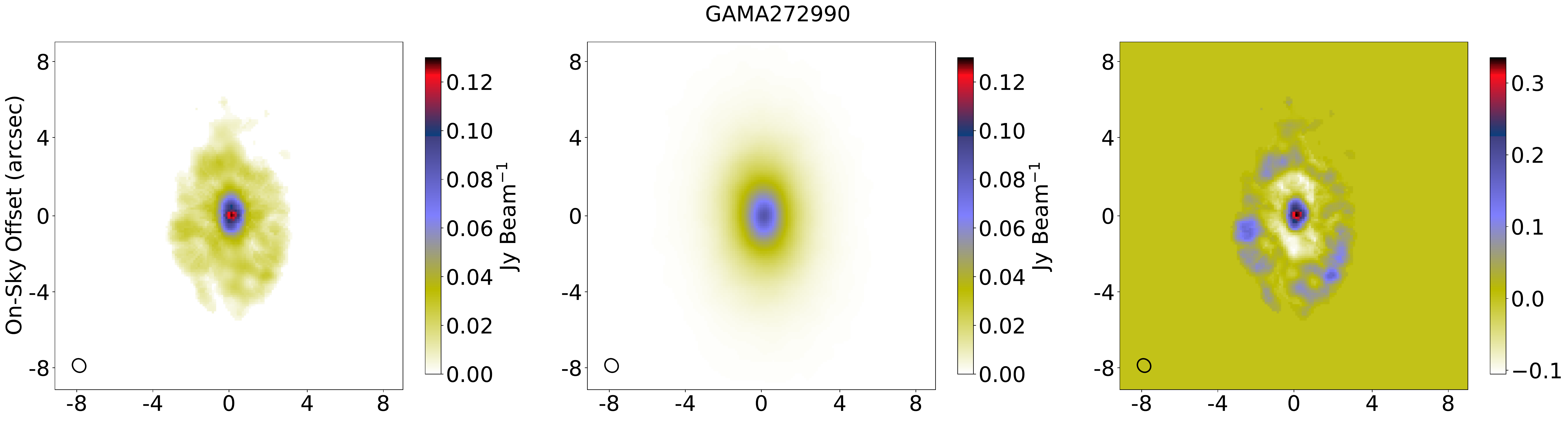}
   \includegraphics[width=\textwidth]{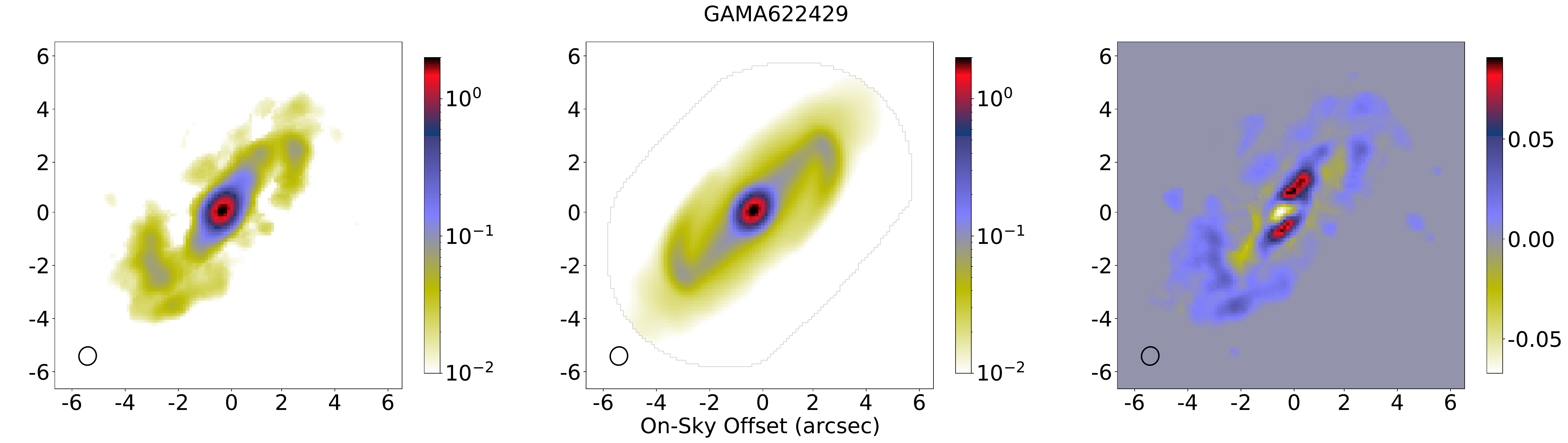}
   \caption{Zeroth order moment maps for (\textit{left}) smoothmasked data, (\textit{centre}) model and (\textit{right}) smoothmasked fractional residual ((data - model)/maximum value of data) for (\textit{top}) GAMA64646, (\textit{middle}) GAMA272990, (\textit{bottom}) GAMA622429. Images for GAMA622429 are log-normalised and truncated at 0.01 Jy Beam$^{-1}$. Synthesised beam FWHM is shown in black. Colours indicate emitted flux in Jy beam$^{-1}$ or residual.}
    \label{fig:Zeroth_Order_Maps}
\end{figure*}
\begin{figure*}
   \includegraphics[width=\textwidth]{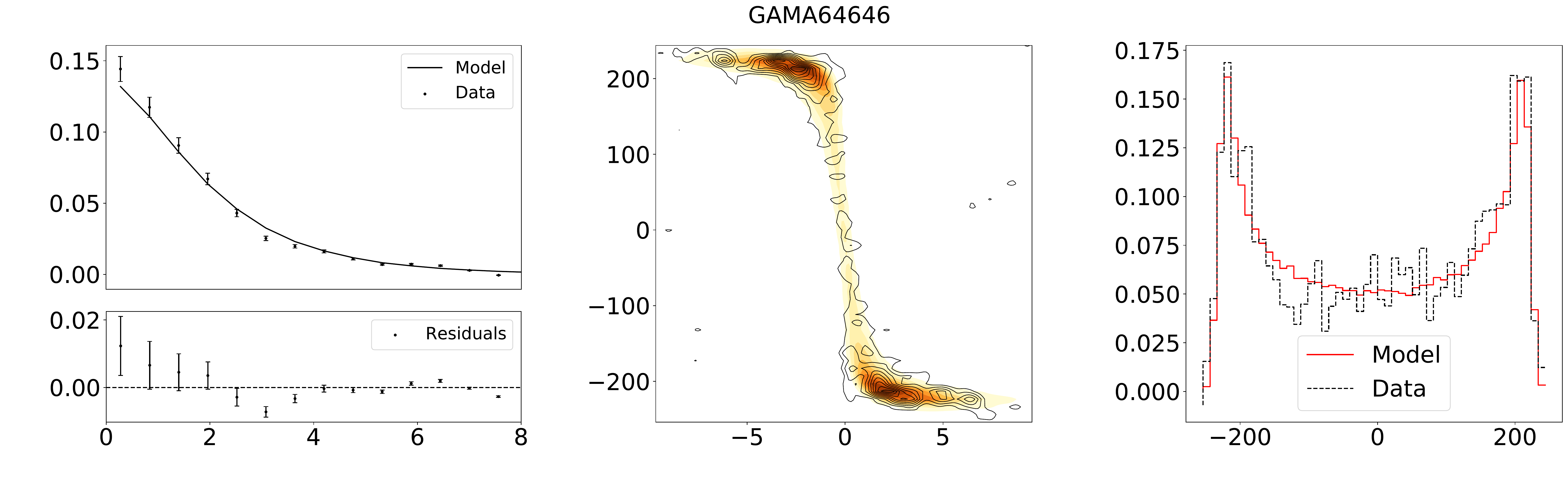}
   \includegraphics[width=\textwidth]{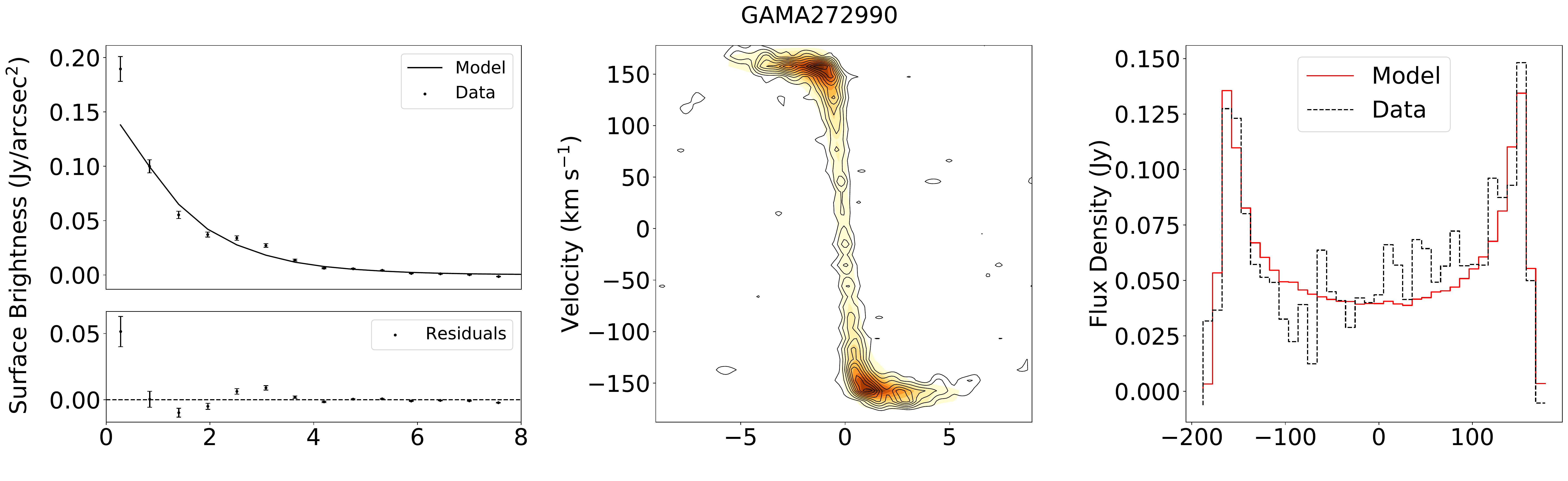}
   \includegraphics[width=\textwidth]{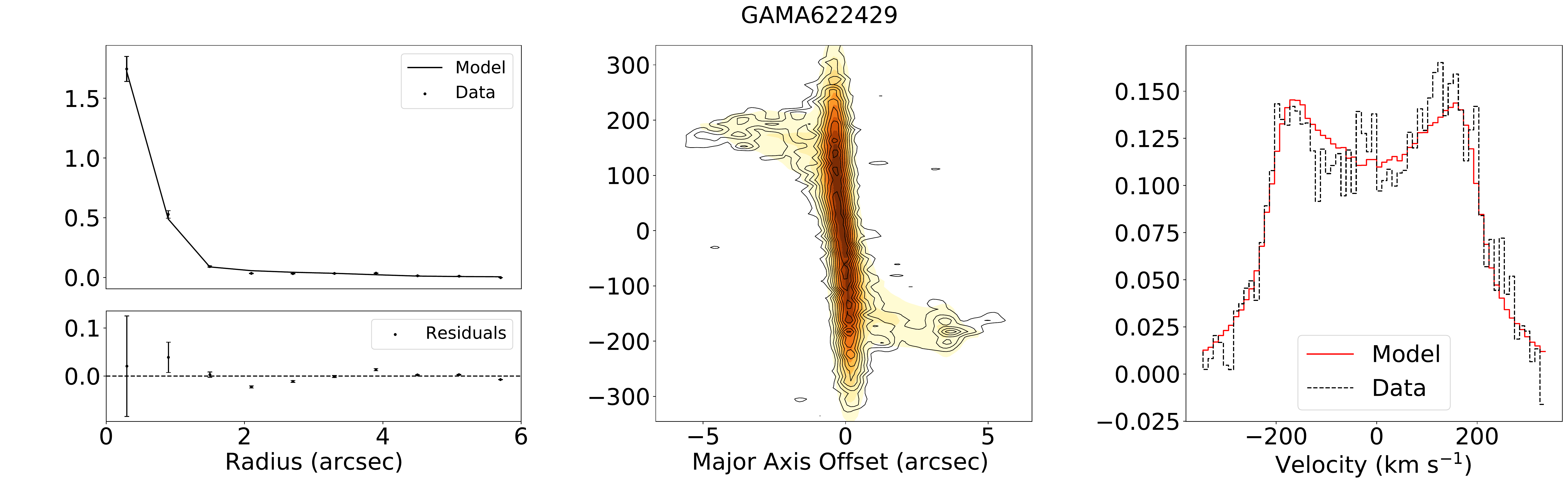}
   \caption{Diagnostic plots for (\textit{top}) GAMA64646, (\textit{middle}) GAMA272990, (\textit{bottom}) GAMA622429. \textit{Left}: Azimuthal average surface brightness profiles. \textit{Centre}: Position-Velocity (PV) Diagrams, data shown as contours, model shown as colour. \textit{Right}: spectra, derived from the total flux per velocity frame within the elliptical masks.}
    \label{fig:Diagnostic_Plots}
\end{figure*}

Figure \ref{fig:Zeroth_Order_Maps} shows zeroth order moment maps for GAMA64646, 272990 and 622429, including ALMA data, fitted models and residuals (data - model). Maps for data and residuals have been smoothmasked \citep{Dame11} to highlight faint emission, using a mask generated for data in both cases. Figure \ref{fig:Diagnostic_Plots} shows azimuthally-averaged surface brightness plots, position-velocity (PV) diagrams for a 1-beam (5-pixel) strip along major axes and spectra for each galaxy. Findings for each galaxy are discussed in turn below.

The ALMA data for GAMA64646 and GAMA272990 were fitted with a model based on a single axisymmetric disc with a S\'ersic surface brightness profile and a single parameter for velocity dispersion. Total fluxes were fixed at the value in Table \ref{tab:Ellipse_Masks}. Table \ref{tab:64646_272990_fit_params} shows the parameter values and 1$\sigma$ uncertainties using MCMC fitting of models to data. 

Attempts were also made to model additional features (central unresolved source and outer ring) using \textsc{KinMS}, introducing six additional parameters. Small reductions ($\sim$1\% and $\sim$3\%) in $\chi^2$ were achieved, with $\chi^2$ approximately equating to Bayesian Information Criteriion \citep[BIC,][]{Schwartz78, Liddle07} because of the very large sample size.  The BIC is given by Equation \ref{eq:BIC}, with $k$ the number of fitted parameters, $n$ the number of data points fitted and $\hat{L}$ the maximised Bayesian likelihood.

\begin{equation}
    BIC = k\ln\left( n \right) - 2 ln \left( \hat{L} \right)
	\label{eq:BIC}
\end{equation}

The very small reduction in BIC indicates a slight preference for the more complex model.  However, attempts to extract posterior distributions for the additional parameters using the increased RMS noise discussed above were not successful, with uncertainties at least as large as the fitted parameters. The same effect was found when attempting to fit an unresolved central source only along with a disc. The single-disc model is therefore discussed below.

\begingroup
\renewcommand{\arraystretch}{1.3} 
\begin{table*}
	\centering
	\caption{Parameters for fitted molecular gas disc models with a S\'ersic surface brightness profile for GAMA64646 and GAMA272990. Spatial and velocity centroids were also fitted but are not shown.}
	\label{tab:64646_272990_fit_params}
	\begin{tabular}{lcccc}
		\hline
             \vspace{-0.1cm}
                          & GAMA64646 & & GAMA272990 & \\
             \vspace{-0.1cm}
		Parameter & Prior Range & Value & Prior Range & Value \\
                            & Lower/Upper &        & Lower/Upper & \\
            \hline
		Position angle ($\degree$)                & 40/90                  & 62.0 $\pm$ 0.5 &  -7/13   &  3.49 $\pm$ 0.81  \\
            Inclination ($\degree$)                      & 45 /85                 & 66.6 $\pm$ 0.9 &  35/75  & 50.9 $\pm$ 1.9 \\
            Max. circular velocity (km s$^{-1}$ ) &  150/350             & 250.9 $\pm$ 3.0 &  100/300  & 211.1 $\pm$ 5.4 \\
            Velocity scaling factor (arcsec)            &  10$^{-6}$/0.5   & 0.18 $\pm$ 0.05 &  10$^{-6}$/1.5  & 0.047 $\pm$ 0.02 \\
            SB profile scale factor (arcsec)            &  0.01/6               & 3.9 $\pm$ 0.5     &  0.01/7  & 1.59 $\pm$ 0.60 \\
            S\'ersic index                                    &  0.05/3               & 0.70 $^{+0.11}_{-0.09}$  &   0.05/4    & 1.02 $^{+0.25}_{-0.20}$ \\
            Velocity dispersion (km s$^{-1}$)      &  1/30                 & 7.1 $^{+1.3}_{-1.1}$  &  1/30 & 6.0 $^{-1.2}_{-1.0}$  \\
		\hline
	\end{tabular}
\end{table*}
\endgroup

\subsubsection{GAMA64646}

Comparison of model and data indicate that the molecular gas disc in GAMA64646 is relatively axisymmetric. Symmetry is apparent in the residual map, the PV diagram and the spectrum (Figures \ref{fig:Zeroth_Order_Maps} and \ref{fig:Diagnostic_Plots}). The arctangent-based rotation velocity profile model accounts well for the profile in the data, apparent in the PV diagram. The residual map and the elliptical azimuthal average plot both indicate that there are weak additional features present. A bright, unresolved central feature creates additional surface brightness at the centre, and bright features away from the centre are suggestive of an inner ring-like structure $\sim$4 arcsec from the centre. An outer ring also appears to be present in the residual map, and is apparent as a slight bump in the azimuthal average surface brightness profile at $\sim$6 arcsec. Perturbations by interaction with another galaxy can generate ring-like features, such as collisional rings or pseudo-rings. However, rings formed by resonance effects (4:1 Lindblad resonance for inner rings) dominate observed rings in galaxies \citep{Buta96}. 

The PV diagram for GAMA64646 shows minor regions of flux at low offset from the centre but high velocity which are not fitted by the single-component model. The central feature appears to be unresolved, which places an upper limit on its diameter (1 beam, $\sim$5 pixels) of $\sim$450 pc. It might be a small concentration of molecular gas around a supermassive black hole (SMBH), consistent with the presence of a weak AGN (Figure \ref{fig:WHAN}). The profiles in Figure \ref{fig:Diagnostic_Plots} show that the emission from the central feature does not dominate the overall molecular gas emission.

The inner ring includes two diametrically opposed bright patches, apparent in the residual zeroth-order map. These could be an effect of observing the inclined ring, where an apparent overlap of the front and rear portions of the ring is viewed along the line of sight. They could also be an effect associated with a bar, where the bar ends interact with the inner ring (see \citet{Buta96} for examples in optical images). The zeroth order moment map for GAMA64646 (Figure \ref{fig:Zeroth_Order_Maps}) shows no indication of a bar. Instead, an annular region partially depleted in molecular gas completely encircling the bright central feature is apparent. A former bar could have been disrupted \citep[e.g.][]{Comeron14} possibly by an interaction, or molecular gas could have been funnelled by a bar towards the centre creating the bright central feature.Overall, GAMA64646 may have undergone a past perturbation (e.g. $>$100 Myr ago, \citet{vandevoort18}), and has settled into its observed configuration.

\subsubsection{GAMA272990}

For GAMA272990, a disc of molecular gas with an asymmetric distribution between North and South is apparent in the zeroth order moment map (Figure \ref{fig:Zeroth_Order_Maps}), with more emission in the lower part of the image than the upper. The spectrum (Figure \ref{fig:Diagnostic_Plots}) shows a noticeable tilt, with flux density increasing from negative to positive velocity, consisent with the spatially asymmetric emission. There is also an irregular emission region at the North of the molecular gas disc, apparent in the zeroth order moment map. The PV diagram (Figure \ref{fig:Diagnostic_Plots}) shows that the model with the arctangent-based circular velocity profile (Equation \ref{eq:velprof}) fits the data well. The azimuthally-averaged surface brightness plot (Figure \ref{fig:Diagnostic_Plots}) shows that the single S\'ersic profile disc model underestimates the emission at the centre of the ETG and at a region $\sim$ 3 arcsec from the centre. The residual zeroth order moment map (Figure \ref{fig:Zeroth_Order_Maps}) shows that there is a brighter central feature and a ring-like structure at $\sim$3 arcsec. However, the central feature appears to be unresolved, and as with GAMA64646 this could be a concentration of molecular gas with an upper limit on radius of 1 beam ($\sim$500pc). GAMA272990 has a weak AGN (Figure \ref{fig:WHAN}), consistent with this finding. The profiles in Figure \ref{fig:Diagnostic_Plots} show that the emission from the central feature does not dominate the overall molecular gas emission. 

The KiDS r-band image (Figure \ref{fig:r_gas_dust_overlay}) shows that GAMA272990 has the optical characteristics of an elliptical galaxy beyond the embedded CO disc (Figure \ref{fig:r_gas_dust_overlay}). This elliptical structure could indicate a past energetic disturbance, randomising previously circular stellar orbits and forming a stellar halo. One possibility for the recent evolution of GAMA272990 is the merger of two gas-rich disc galaxies, forming an elliptical with an embedded gas disc with the combined rotational energy of the gas from both progenitors \citep{Ueda14}. It is also possible that the disc is the result of a merger between an existing elliptical galaxy and a gas-rich object.  It is likely that GAMA272990 was disturbed relatively recently, causing molecular gas to settle into the observed disc/ring structure and creating a lop-sided distribution of molecular gas. Molecular gas has accumulated around the centre of the ETG, forming the bright unresolved central region which could be associated with a central SMBH. 

\subsubsection{GAMA622429}
\label{sec:622429}

\begin{figure}
   \includegraphics[width=\columnwidth]{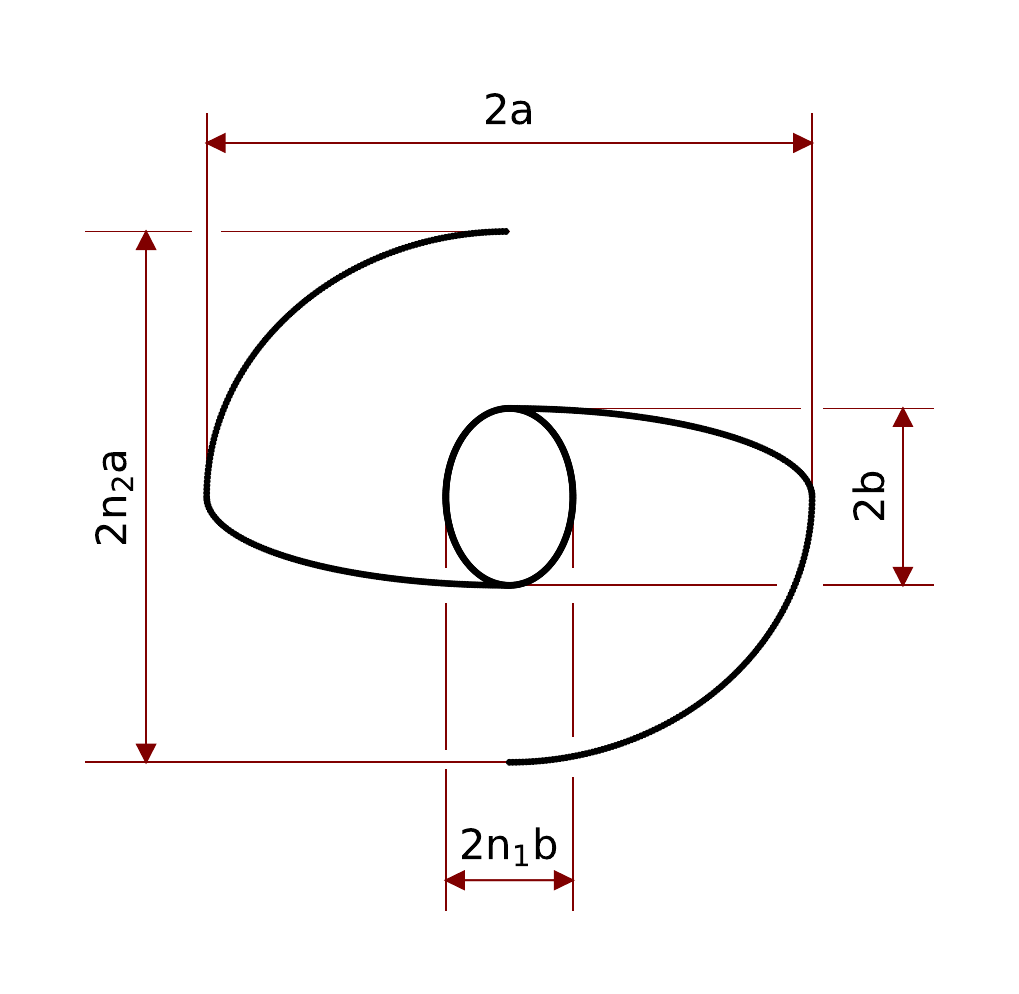}
   \caption{Plan view of point-source model for nuclear ring, bar and spiral arms using elliptical pathways. All dimensons are defined by the bar major and minor radii, a and b, with scaling parameters n$_1$ and n$_2$.}
    \label{fig:Ring_Bar_Arms_Params}
\end{figure}

Initial attempts at kinematic modelling for GAMA622429 using a single disc with a S\'ersic surface brightness profile were not successful, with the the model fitting a bright bar-like feature as a disc viewed edge-on. This model is not considered to be representative of features within the ALMA observation, due to the unlikely edge-on orientation relative to the host galaxy. The zeroth order moment map for GAMA622429 (Figure \ref{fig:Zeroth_Order_Maps}) shows a bright central feature resembing a nuclear ring, a bar and faint spiral arm structure (Figure \ref{fig:r_gas_dust_overlay}). These were modelled by positioning point sources to create these structures, using the \textsc{inClouds} functionality in \textsc{KinMSpy}. Point sources were distributed along loci defined by ellipses, with a whole ellipse perpendicular to the bar used for a nuclear ring and opposite quadrants of ellipses used for a bar and spiral arms. \citet{Davis13} show examples of models of bar and spiral structures (based on logarithmic spiral loci) in their Figure 10. \citet[][their sections 4, 5 and 6]{Sellwood93} provide further information on galactic bars, their associated stellar orbits and their relationship to observed molecular gas distributions modelled by \citet{Athanassoula92} which form the basis of the point source model in Figure \ref{fig:Ring_Bar_Arms_Params}. The bar major radius a and minor radius b define all these features, with b also used for the nuclear ring major axis and a for the spiral arm major axis. Separate factors n$_1$ and n$_2$ were applied to a and b to define the minor axis for the elliptical loci of the nuclear ring and the spiral arm. The value for n$_2$ was fixed at 1.72 as indicated by initial model fitting using scipy.optimize.minimize, because initial MCMC-based model fitting shows this parameter to be highly uncertain. Point sources were distributed uniformly along their loci, using an empirical model to position sources equally around ellipses or elliptical segments. These additional components were modelled with 100,000 point sources in total, distributed amongst the three features in proportion to their emitted flux. Bisymmetric flow was applied to the cloudlets within the bar region using the description of \citet{Spekkens07} as implemented in \textsc{KinMSpy}, defined by radial and tangential velocities in relation to the galaxy centre and the a and b parameters for scaling. This was superimposed on an axisymmetric radial velocity profile as described in Section \ref{sec:Kinematic_Modelling}. A fixed position angle of 6$\degree$ was also applied to the nuclear ring component, as indicated by initial fitting but with high uncertainty.

\begin{figure}
   \includegraphics[width=\columnwidth]{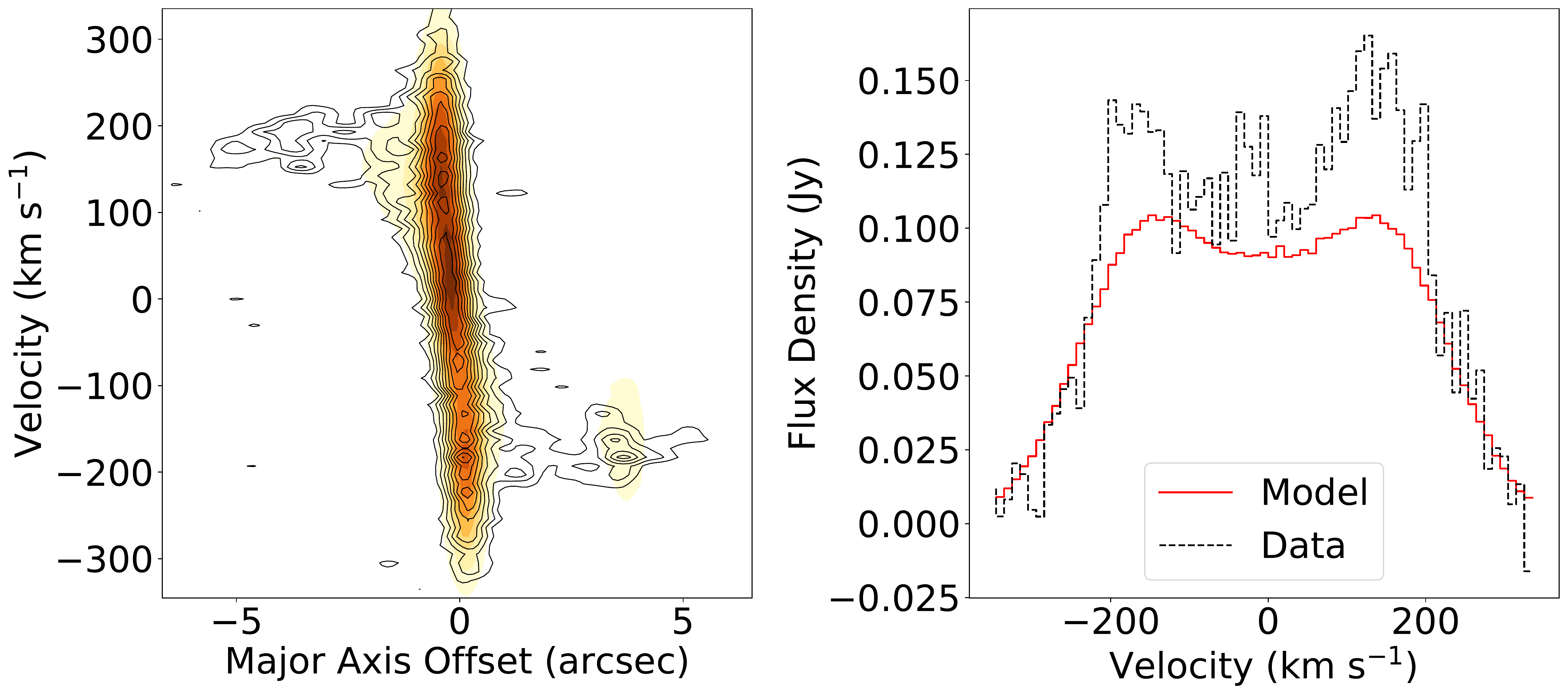}
   \caption{PV diagrams (data as contours, model in colour) and spectra for GAMA622429 fitted with a ring + bar + spiral arms model only.}
    \label{fig:G622429_Clouds_Only}
\end{figure}

The resultant fitted model on its own (using \textsc{scipy.optimize.minimize}) was only partially successful, leaving an unfitted double-horned feature in the spectrum suggestive of an underlying rotating disc (Figure \ref{fig:G622429_Clouds_Only}). The position angle (PA) of the fitted structure was found to be $\sim$135$\degree$, consistent with the PA of the overall molecular gas distribution for this ETG. Figure \ref{fig:Zeroth_Order_Maps} confirms that the molecular gas bar and the major axis of the overall molecular gas distribution are fortuitously aligned. A rotating molecular gas disc with a S\'ersic surface brightness profile and a common PA for all features was therefore added to the model, using the approach described in Section \ref{sec:Kinematic_Modelling}. Attempts to fit a separate PA for the bar compared to the disc will lead to high uncertainty in the result because of the high inclination of the ETG.

Table \ref{tab:622429_fit_params} shows the parameter values and 1$\sigma$ uncertainties using MCMC fitting of this refined model to the data. The PV diagram for data and model (Figure \ref{fig:Diagnostic_Plots}) shows that this model fits the data well. The zeroth order moment map for residuals (Figure \ref{fig:Zeroth_Order_Maps}) shows a slight decline in bar surface brightness with radius compared to the flat profile of the model, similar to profiles in optical images for bars in SB0 galaxies \citep[][and references therein]{Sellwood93}. This is also reflected in the model for the nuclear ring, which is elongated along the length of the bar (n$_1$ = 2.2). This effect has probably led to an over-estimate of the flux allocated to the nuclear ring model, by including flux from the inner regions of the bar. The total flux of the model is 64.0$^{+5.9}_{-5.7}$ Jy km s$^{-1}$, which agrees with the value from curve of growth analysis within the uncertainties (Table \ref{tab:Ellipse_Masks}). Almost half of this emission arises from the nuclear region of the ETG.

\begingroup
\renewcommand{\arraystretch}{1.3} 
\begin{table}
	\centering
	\caption{Parameters for a molecular gas central ring, bar and spiral arm structure plus an underlying disc with a S\'ersic surface brightness profile for GAMA622429.}
	\label{tab:622429_fit_params}
	\begin{tabular}{lcc}
		\hline
             \vspace{-0.1cm}
		Common Parameters & Prior Range & Value  \\
                                            & Lower/Upper &        \\
            \hline
		Position angle ($\degree$)                       & 100/170               & 136.8 $\pm$ 1.5 \\
                     Inclination ($\degree$)                             & 50/85                   &  72.3 $^{+2.3}_{-2.9}$ \\
                     Far-field circular velocity (km s$^{-1}$ )  & 160/260               & 207.8 $^{+9.1}_{-8.2}$  \\
                     Vel. profie scale factor (arcsec)                & 0.01/1                  & 0.12 $^{+0.15}_{-0.08}$ \\
		\hline
             \vspace{-0.1cm}
		Parameter (Nuclear ring, bar, spiral arms) & Prior Range & Value  \\
                                                                         & Lower/Upper &         \\
            \hline
		        Line flux (nuclear ring) (Jy km s$^{-1}$)  & 1/40                     & 28.5 $^{+1.6}_{-1.8}$\\
                    Line flux (bar) (Jy km s$^{-1}$)                & 0.01/50                & 3.5 $^{+2.8}_{-2.3}$ \\
                    Line flux (spiral arms) (Jy km s$^{-1}$)    & 1/16                      & 3.4$^{+1.8}_{-1.5}$ \\
                    Velocity dispersion (km s$^{-1}$)              & 30/170                 & 99.1 $^{+11.8}_{-16.8}$ \\
                    Bar radius (a) (arcsec)                                & 2/6                      & 3.8 $^{+0.4}_{-0.3}$ \\
                    Nuclear ring major radius (b) (arcsec)         & 0.001/1                 & 0.23 $\pm$ 0.07  \\
                    Nuclear ring minor radius scale factor (n$_1$)        & 0.001/6                  & 2.2 $^{+0.7}_{-0.9}$ \\
                    Bar radial velocity (km/s)                             & 0.1/350                 & 79.2 $^{+55.0}_{-34.3}$ \\
                    Bar tangential velocity (km/s)                     & 0.001/150             & 48.8 $^{+43.4}_{-29.8}$ \\
		\hline
             \vspace{-0.1cm}
		Parameter (Disc) & Prior Range & Value  \\
                                     & Lower/Upper &          \\
            \hline
            Flux (Jy km/s)                                                     & 1/50                       & 28.7  $\pm$ 4.6 \\
            SB profile scale factor (arcsec)                           & 0.001/15               & 1.7 $^{+1.5}_{-1.1}$ \\
            S\'ersic index                                                       & 0.1/4                     & 1.38 $^{+0.60}_{-0.49}$  \\
            Velocity dispersion (km s$^{-1}$)                       & 1/50                      & 25.9 $^{+7.3}_{-6.1}$ \\
		\hline
	\end{tabular}
\end{table}
\endgroup

The fitted spectrum for GAMA622429 (Figure \ref{fig:Diagnostic_Plots}) shows some excess flux, mainly at positive velocity but also as a spike at relative velocities just below zero. The latter does not correlate to any features and may be a noise effect. The residual zeroth order moment map (Figure \ref{fig:Zeroth_Order_Maps}) also shows excess flux, as a diffuse extended structure coinciding with the spiral arm in the lower portion of the image (positive velocity). One explanation for this is recent minor merger activity, which has added molecular gas to the ETG with insufficient time to allow it to smooth out fully. The KiDS optical image (Figure \ref{fig:r_gas_dust_overlay}) supports this, with irregularities apparent in the lower portion of the disc where the diffuse patch of excess CO emission was detected. Molecular gas could then have been funnelled towards the galaxy centre, forming the very bright central feature and fuelling its strong AGN (Figure \ref{fig:WHAN}). Alternatively, the action of the bar could be disturbing molecular gas within the spiral arms on that side of the galaxy. The bar itself could have been produced following a perturbation.


\subsection{Molecular Gas Stability Analysis}
\label{sec:Toomre}


The possibility for new star formation within the ETGs with dominant molecular gas discs (GAMA64646 and 272990) was explored using the Toomre stability criterion, Q \citep{Toomre64, Boizelle17, vandevoort18}, which estimates whether a perturbation within a rotating gas disc subject to velocity dispersion would lead to gravitational collapse and the formation of star-forming regions. Equations \ref{eq:Toomre1}, \ref{eq:Toomre2} and \ref{eq:Toomre3} summarise the approach, which require deprojected values for gas surface density based on the model galaxy inclination angle. The fitted rotation velocity profiles (Equation \ref{eq:velprof}) and constant velocity dispersion values were used for this. SI units are used throughout. The Toomre stability criterion Q is given by:

\begin{equation}
    Q=\frac{\kappa\sigma_{gas}}{\pi G \Sigma_{gas}}
	\label{eq:Toomre1}
\end{equation}

\begin{equation}
    \kappa=\left(R \frac{d\Omega^2}{dR} + 4\Omega^2 \right)^{0.5}
	\label{eq:Toomre2}
\end{equation}

where $\kappa$ is the epicyclic frequency, $\sigma_{gas}$ is the gas velocity dispersion, $G$ is the gravitational constant, $\Sigma_{gas}$ is the molecular gas surface density, and $\Omega$ is the angular velocity at radius $R$ ($\Omega=v_c/R$). The derivative term in Equation \ref{eq:Toomre2} is evaluated for the arctangent-based profile in Equation \ref{eq:velprof} as follows:

\begin{equation}
    \frac{d\Omega^2}{dR}=\frac{2}{R^3}\left( \frac{2v_{flat}}{\pi} \right)^2\arctan\left(\frac{R}{R_0}\right)\left( \frac{RR_0}{R^2+{R_0}^2}-\arctan\left( \frac{R}{R_0} \right) \right)
	\label{eq:Toomre3}
\end{equation}

Values of Q less than 1 normally show that the molecular gas is unstable against collapse, and so should form stars efficiently. However, the uncertainties in molecular gas mass estimates from $^{12}$CO[2-1] emission are such that this is not an appropriate threshold for this work. The uncertainty of CO to H$_2$ mass-to-light ratio ($\pm$30\%) and $^{12}$CO[2-1] to $^{12}$CO(1-0) line strength ratio ($\pm$50\%) (S19 and references therein) suggest that values of Q $>$2 are indicative of stability against star formation and Q $<$0.5 are indicative of instability leading to star formation. Values of  Q between these limits indicate increasing likelihood of star formation as Q decreases.

Maps of star fomation rate (SFR) per spaxel are available from SAMI DR3. These were used to indicate where star fomation is actually occurring compared to expectations from values of Q. Overall star formation rates are also available for each galaxy from GAMA DR3 in the catalogue MagPhysv06, derived using spectral energy distribution (SED) fitting with MAGPHYS\footnote{http://www.iap.fr/magphys/} \citep{daCunha08}. Table \ref{tab:SFRs} shows the total estimated star formation rates from the SAMI DR3 maps compared to estimates from GAMA DR3. Uncertainties for the SFR estimates from SAMI are derived from the uncertainty in the empirical relation used to estimate them from H$_{\alpha}$ line luminosities \citep{Dopita05,Medling18}.

\begin{figure*}
   \includegraphics[width=.51\textwidth]{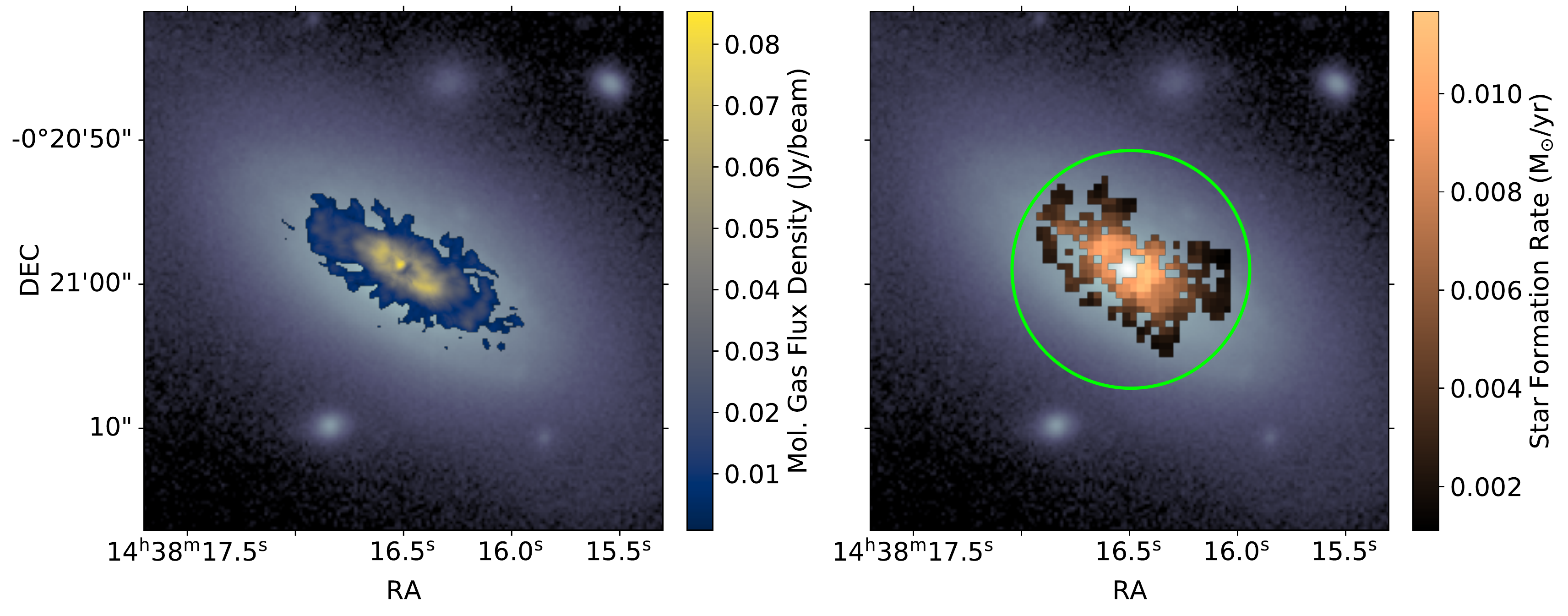}\hfill
   \includegraphics[width=.255\textwidth]{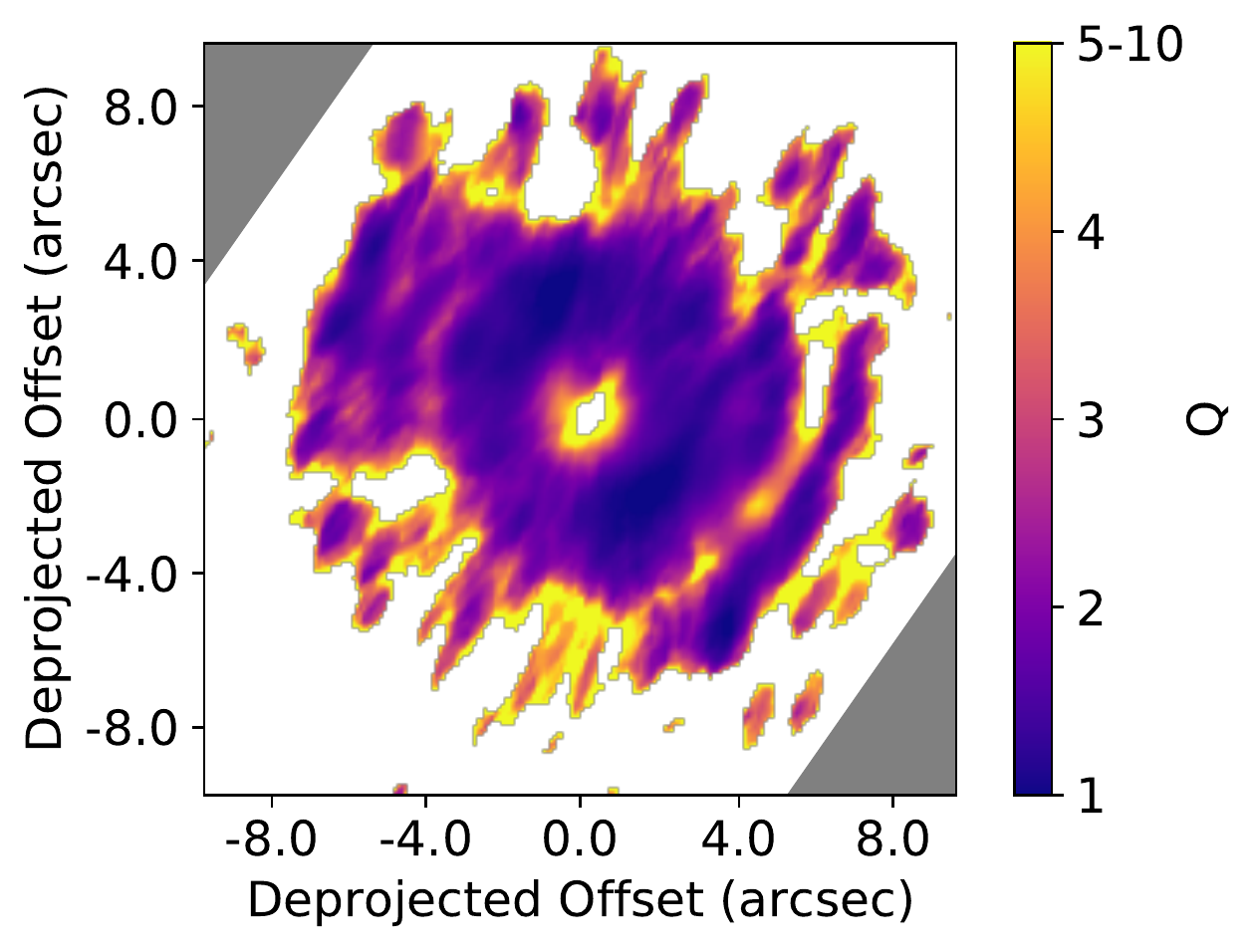}\hfill
   \includegraphics[width=.205\textwidth]{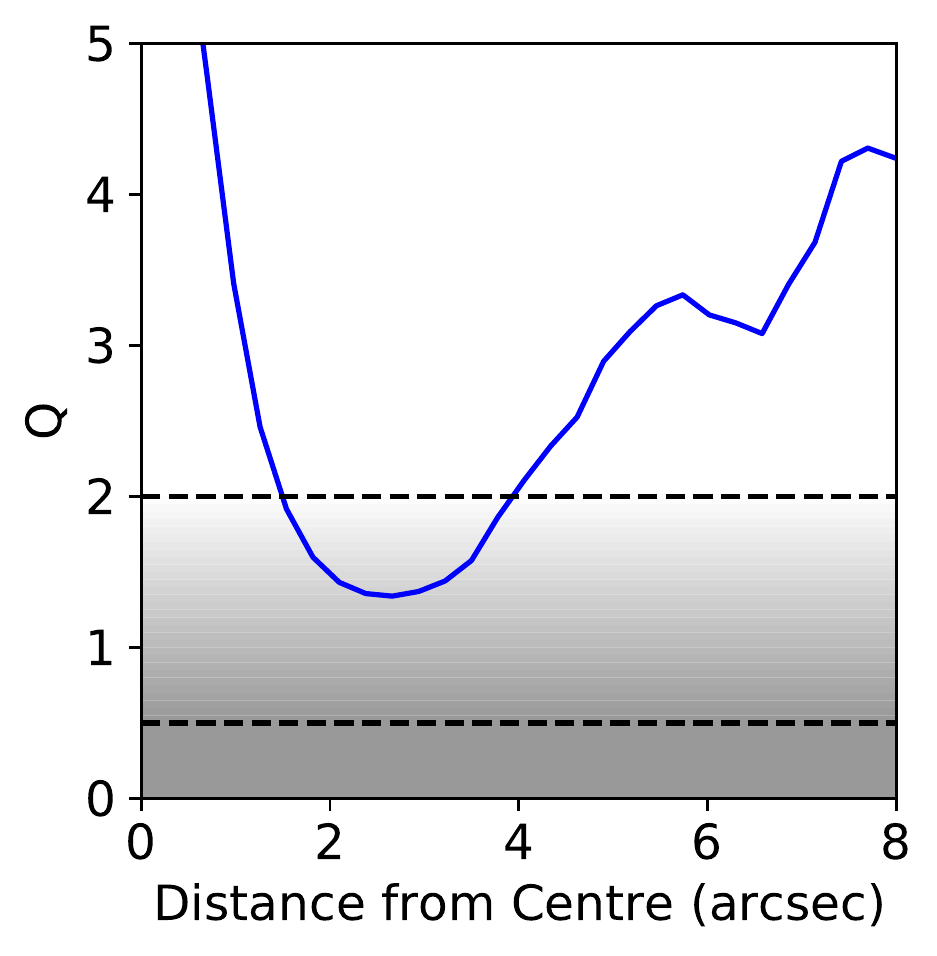}\hfill
   \includegraphics[width=.51\textwidth]{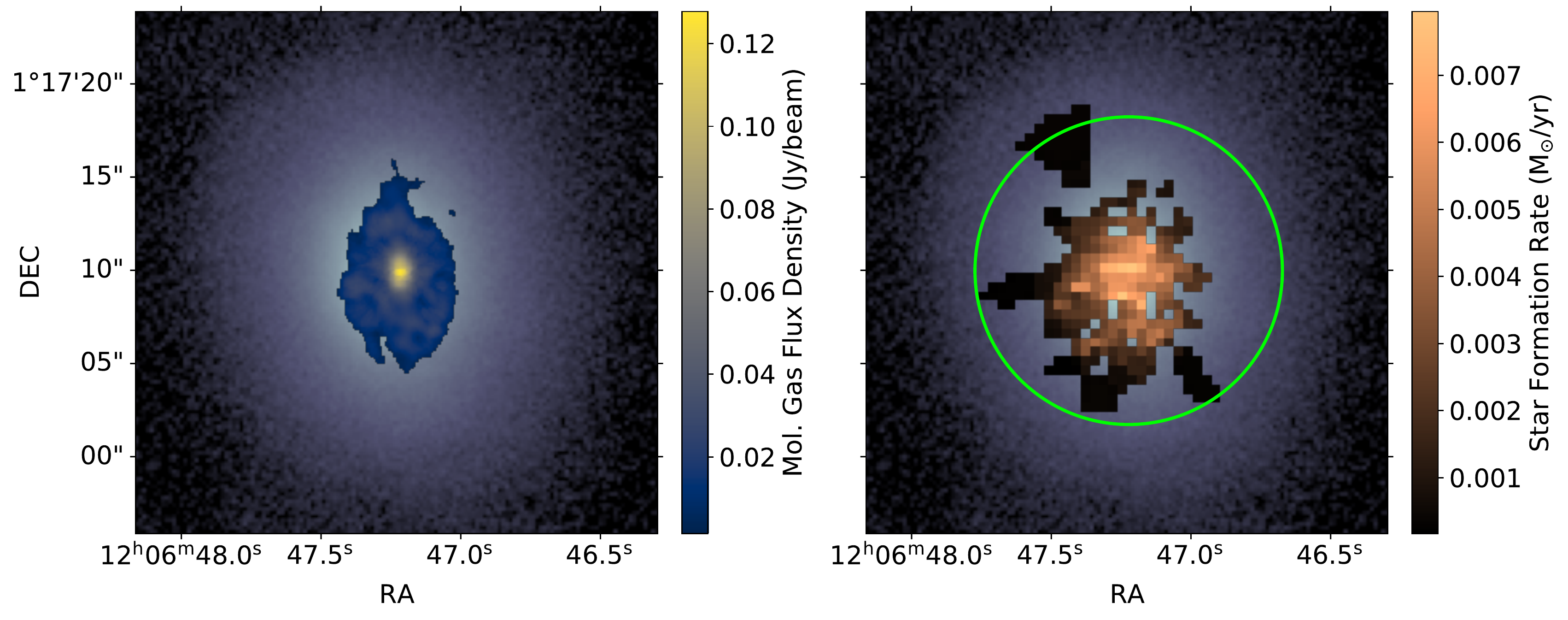}\hfill
   \includegraphics[width=.255\textwidth]{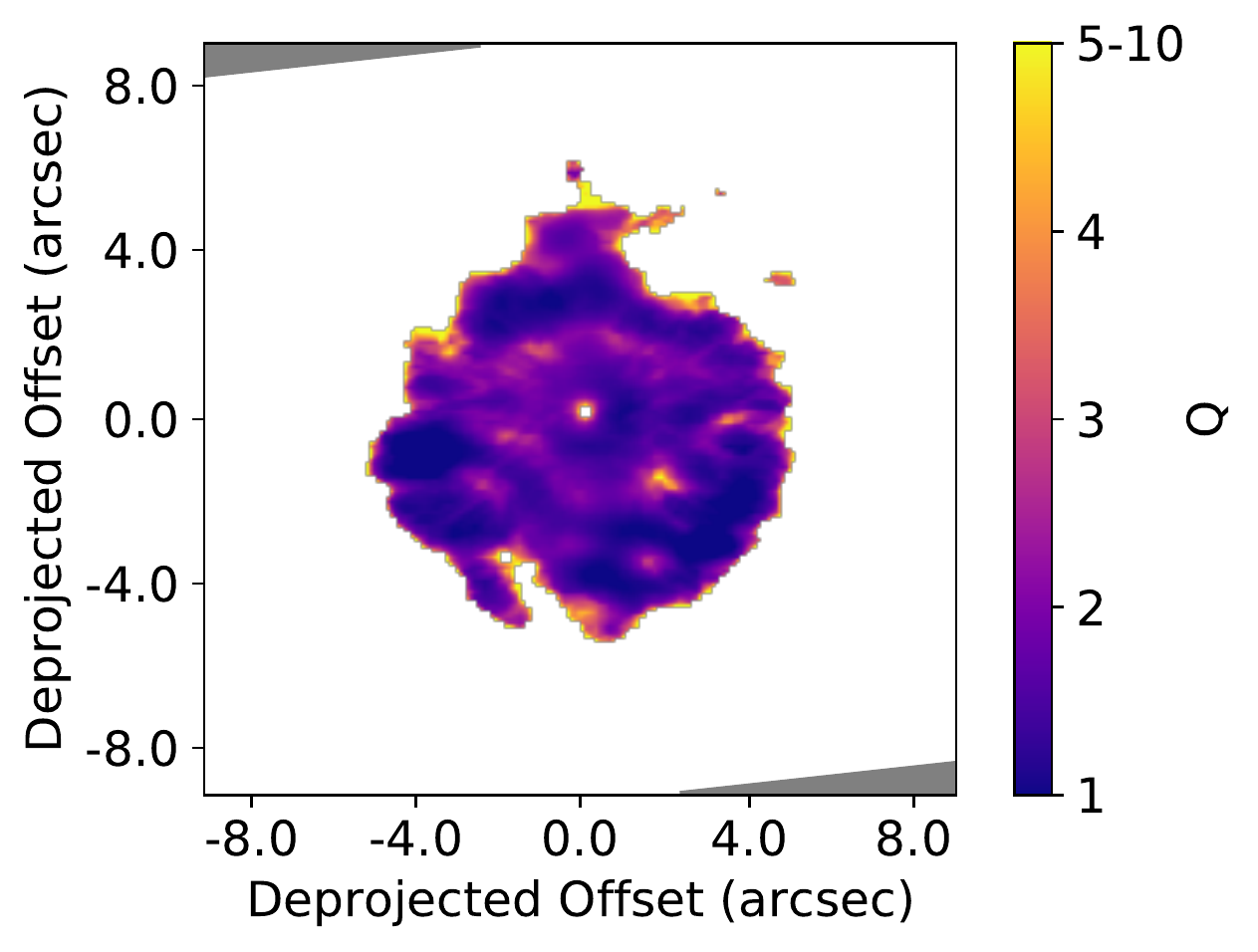}\hfill
   \includegraphics[width=.205\textwidth]{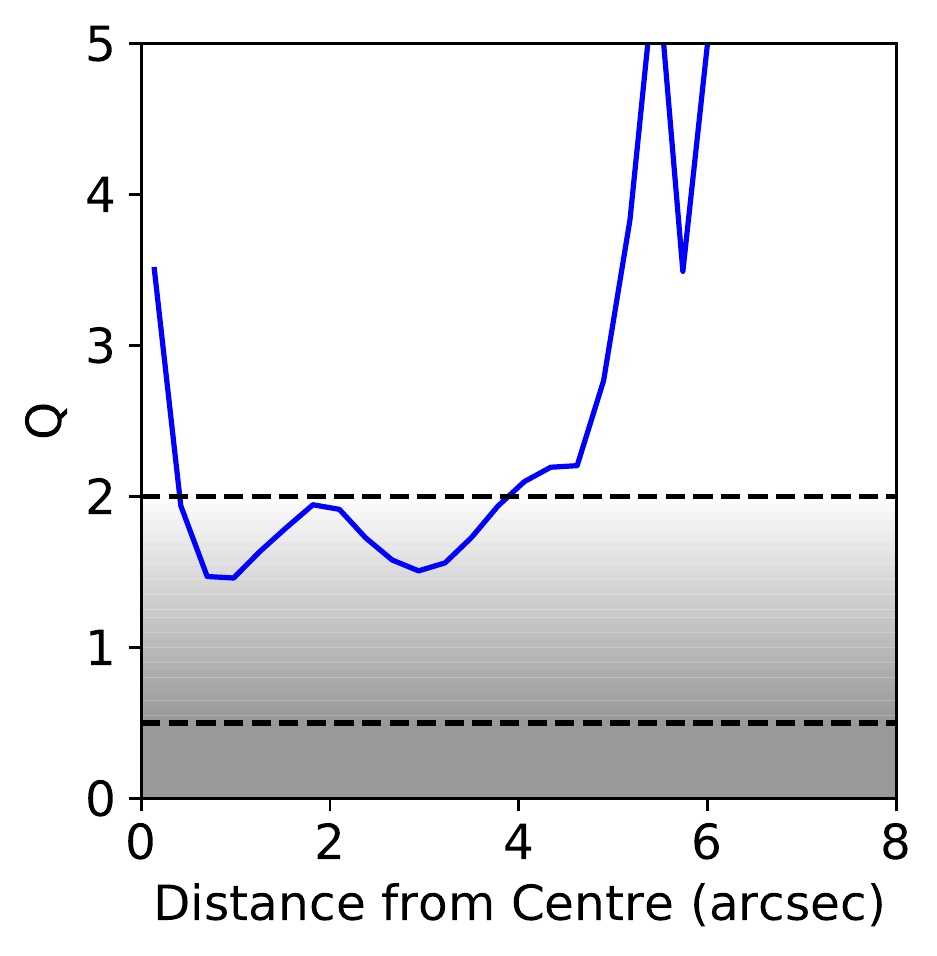}\hfill
   \includegraphics[width=.51\textwidth]{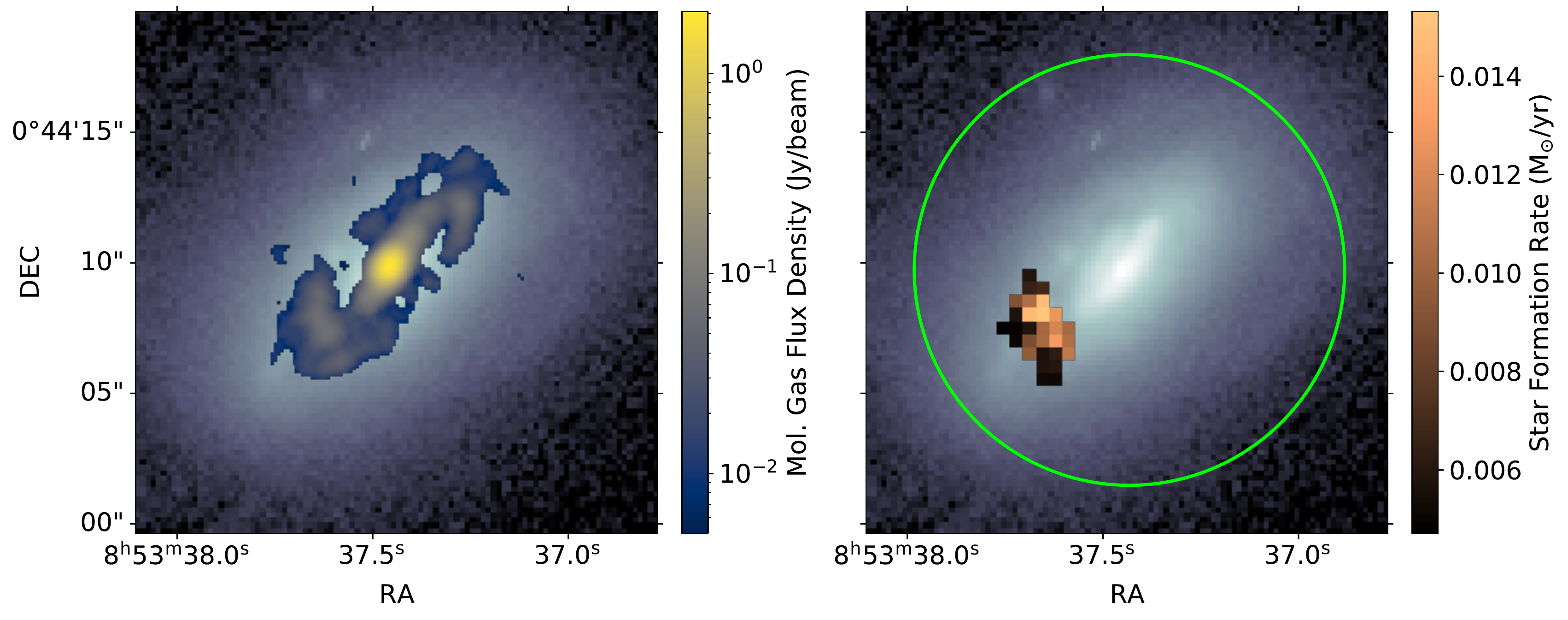}\hfill
   \caption{Comparison of molecular gas distribution, star formation distribution (from SAMI, adaptive binning) and Q for \textit{top:} GAMA64646, \textit{middle:} GAMA272990, \textit{bottom:} GAMA622429. \textit{Left:} smoothmasked zeroth order moment maps of molecular gas overlain on KiDS r-band log-normalised images. Molecular gas distrbution for GAMA622429 is log-normalised. Images are fully contained within the ALMA field of view ($\sim$45 arcsec). \textit{Centre left:} SAMI star formation maps overlain on log-normalised KiDS r-band images. Field of view (16.5 arcsec, \citet{Sharp15}) shown in green. \textit{Centre right:} Maps of Q (deprojected) calculated using constant velocity dispersion and smoothmasked flux density. \textit{Right:} azimuthal average plots of Q, with potential for star formation shaded from white (highly unlikely, Q$>$2) to dark grey (highly likely, Q$<$0.5).}
    \label{fig:SFR_Q}
\end{figure*}

\begin{table}
	\centering
	\caption{Total star formation rates from GAMA DR3 MAGPHYS results (0-100 Myr) and SAMI DR3 (total for maps) for GAMA64646, 272990 and 622429.}
	\label{tab:SFRs}
	\begin{tabular}{lcc}
		\hline
		Galaxy & Star Formation Rate & Star Formation Rate   \\
                       & (GAMA DR3) & (SAMI DR3)  \\
                       & (M$_{\odot}$ yr$^{-1}$) & (M$_{\odot}$ yr$^{-1}$) \\
            \hline
            GAMA64646   & 1.06$^{+0.14}_{-0.00}$ & 1.36 $\pm$  0.09 \\
                                 &                                       &                                           \\
            GAMA272990 &  0.78$^{+0.11}_{-0.17}$ & 0.75 $\pm$ 0.05 \\
                                 &                                       &                                           \\
            GAMA622429 &  4.39$^{+0.59}_{-0.20}$ & 0.25 $\pm$  0.02 \\
	\end{tabular}
\end{table}

Table \ref{tab:SFRs} shows that total SFRs from GAMA DR3 and SAMI DR3 are in reasonable agreement for GAMA64646 and 272990. However, there is a significant discrepancy between values for GAMA622429. This is attributed to the approach taken by the SAMI survey to mask regions where $H_{\alpha}$ emission could be contaminated by emission sources other than star formation \citep{Medling18}. The resultant maps are acknowledged as showing clean (definitely star-forming) but not complete (not all fully identified) areas of a galaxy with star formation. The presence of a strong AGN within GAMA622429 may be causing significant masking of spaxels with star formation. As an indication of which of the two SFR estimates may be correct, the molcular gas depletion times for an approximate molecular gas mass of 7 $\times$ 10$^{9}$ M$_{\odot}$ (S19) are predicted to be $\sim$2 Gyr for the GAMA estimate and $\sim$30 Gyr for the estimate from SAMI. The true estimate of SFR is likely to be between these values, probably closer to the GAMA estimates. Molecular gas depletion times for GAMA64646 and 272990, each containing $\sim$few $\times$ 10$^9$ M$_{\odot}$ (S19) are $\sim$1 Gyr and $\sim$3 Gyr.

Figure \ref{fig:SFR_Q} compares the location of molecular gas in GAMA64646, 272990 and 622429 and the location of star forming regions detected by SAMI. These are affected by atmospheric seeing (FWHM $\sim$2 arcsec, \citet{Croom21}), compared to the ALMA image beam size FWHM of $\sim$0.7 arcsec. The comparison is generally good for GAMA64646 and 272990, but poor for GAMA622429 because of the spaxel masking issues discussed previously.  Fine details such as the absence of star formation at the centre of GAMA64646 are apparent in both the molecular gas map and the star formation map. The SAMI field of view ($\sim$16.5 arcsec) is also shown in the diagram, to highlight that alignment of molecular gas and star formation is not affected by image truncation. This field was derived from \citet{Sharp15}, their Figure 9 which shows how dithered images are stacked to create the field of view. The ALMA field of view ($\sim$45 arcsec) is greater than the size of the images shown.

Figure \ref{fig:SFR_Q} also shows maps of the Toomre stability criterion (Q), and azimuthal average values of Q determined from the maps. Star formation for GAMA64646 is likely in the range 1 - 4 arcsec from the centre, which is in reasonable agreement with the star formation map in Figure \ref{fig:SFR_Q}. The absence of star formation towards the centre is predicted, although the masking of star formation by AGN activity in the SAMI SFR map could be contributing to this. Star formation for GAMA272990 is likely from close to the centre to $\sim$4 arcsec from the centre, which again is in reasonable agreement with actual star formation shown in Figure \ref{fig:SFR_Q}. The equivalent plots for GAMA622429 were not developed because of the presence of the bar, for which the assumption of circular motion of the molecular gas used to calculate Q presented previously is not applicable. Overall, these calculations of Q from the ALMA observations and kinematic modelling successfully predict the observed star formation distribution in these detected massive molecular gas discs.


\subsection{Kinematic Alignments of Molecular Gas, Ionised Gas and Stars}
\label{sec:Kinematic_Alignments}


The relative kinematic PAs of ETGs can be used to indicate whether recent merger activity has occurred. If kinematic PAs of stars and molecular gas are well aligned, a significant merger is unlikely to have deposited additional ISM into the ETG recently e.g. within a few dynamical timescales ($\sim$100 Myr). Conversely, poor alignment (or even counter-rotation) can indicate relatively recent deposition of ISM via an external source \citep{Davis11}. 

\begin{figure*}
   \includegraphics[width=.333\textwidth]{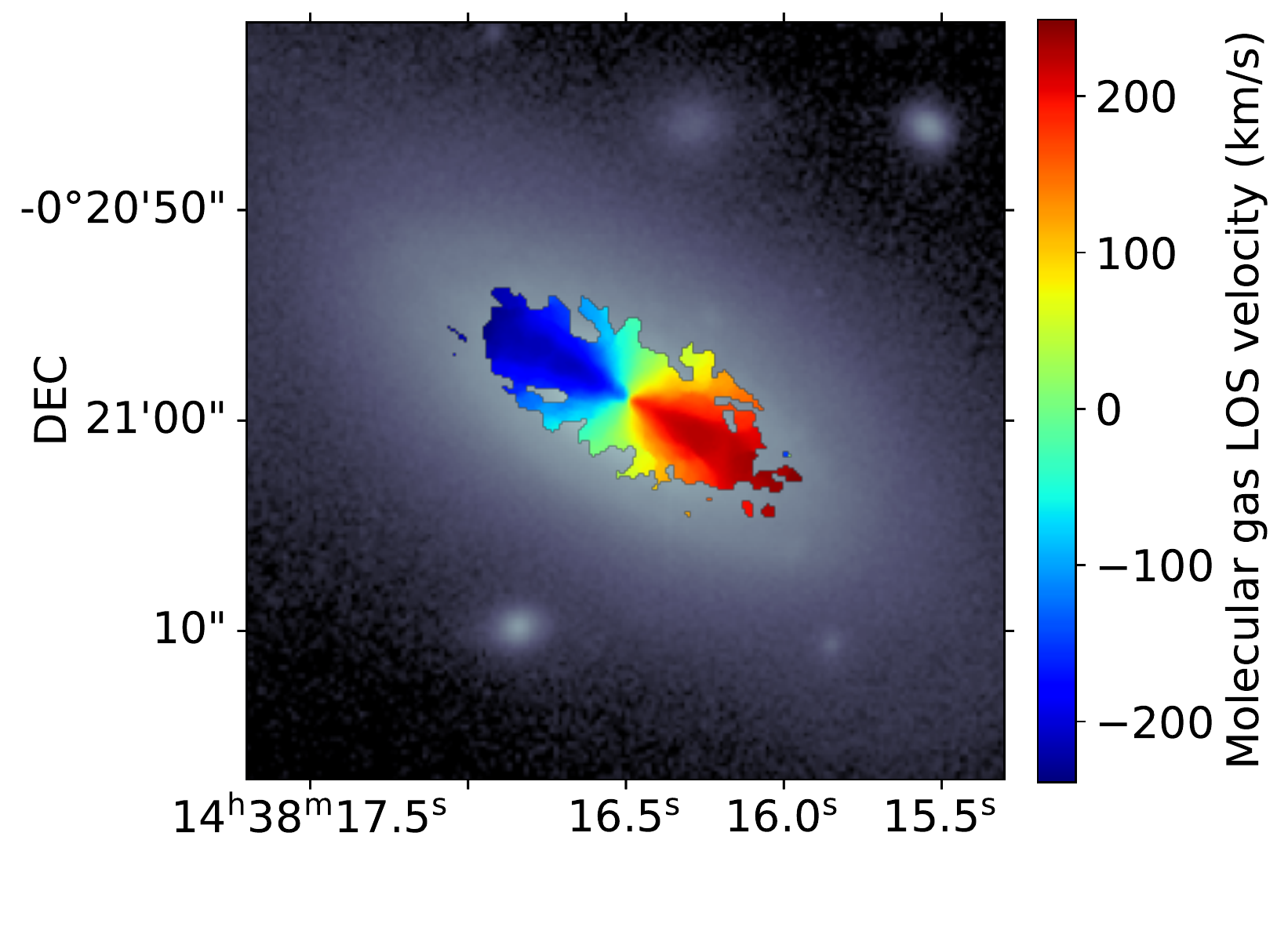}\hfill
   \includegraphics[width=.333\textwidth]{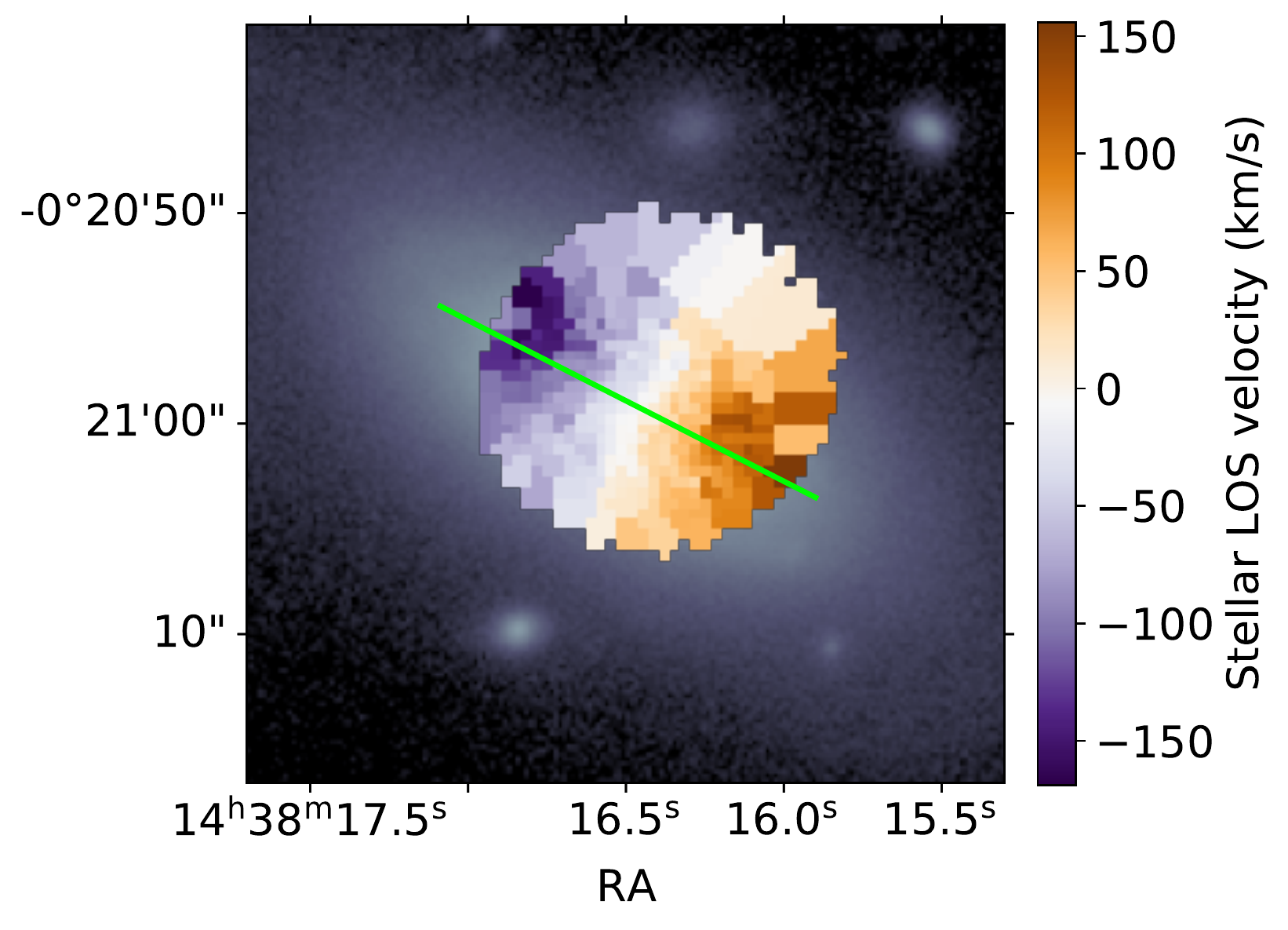}\hfill
   \includegraphics[width=.333\textwidth]{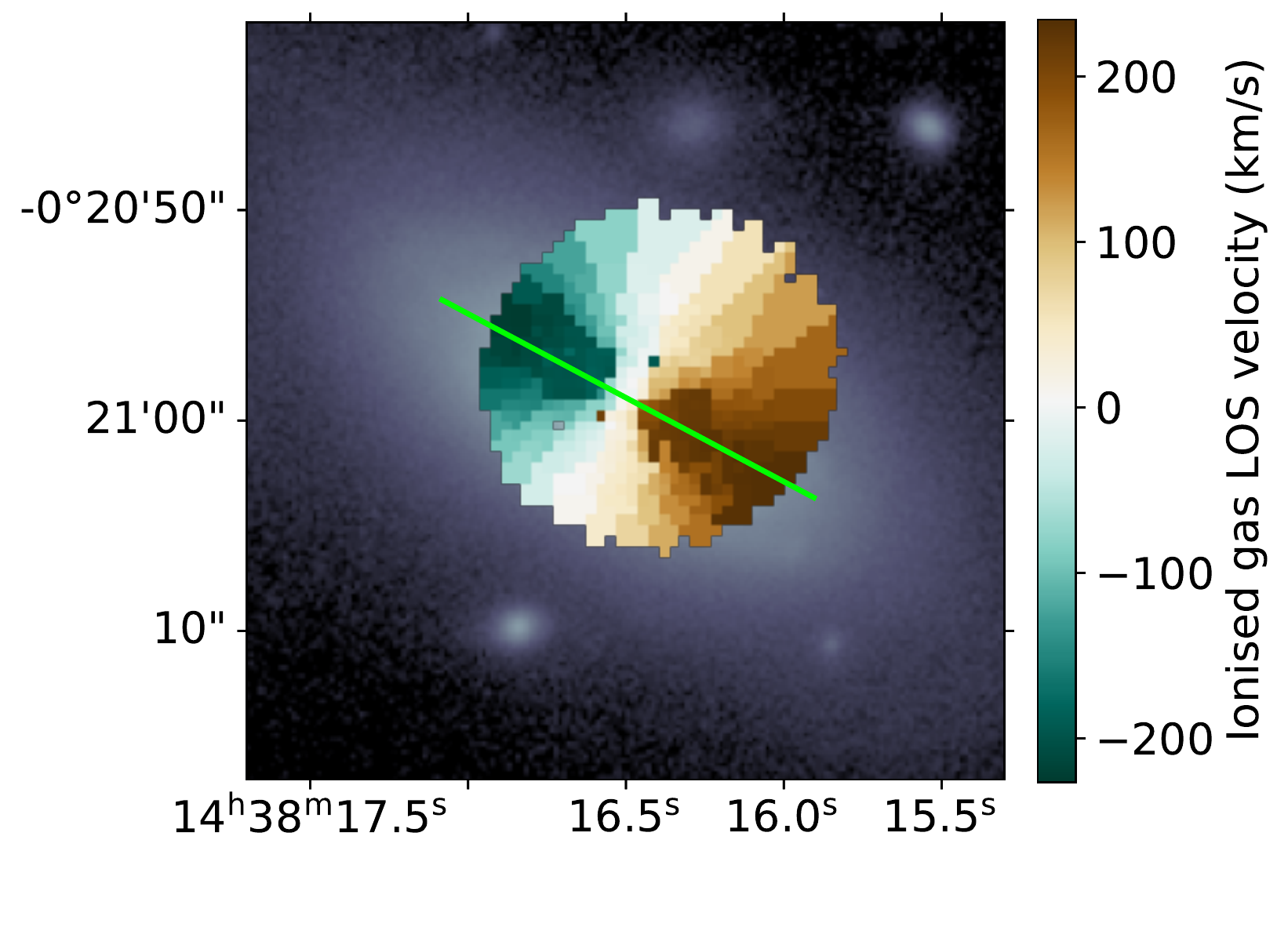}\hfill
   \includegraphics[width=.333\textwidth]{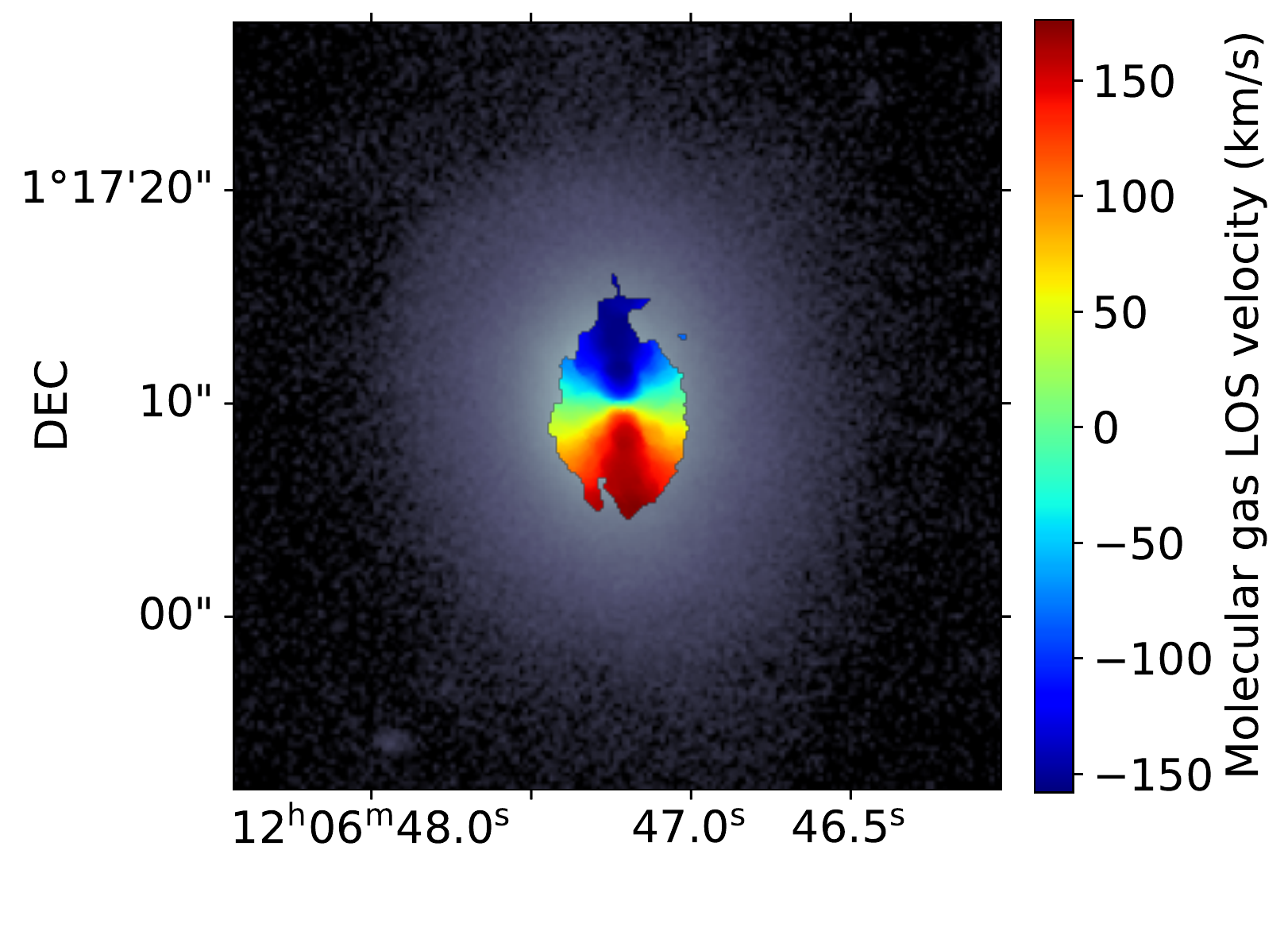}\hfill
   \includegraphics[width=.333\textwidth]{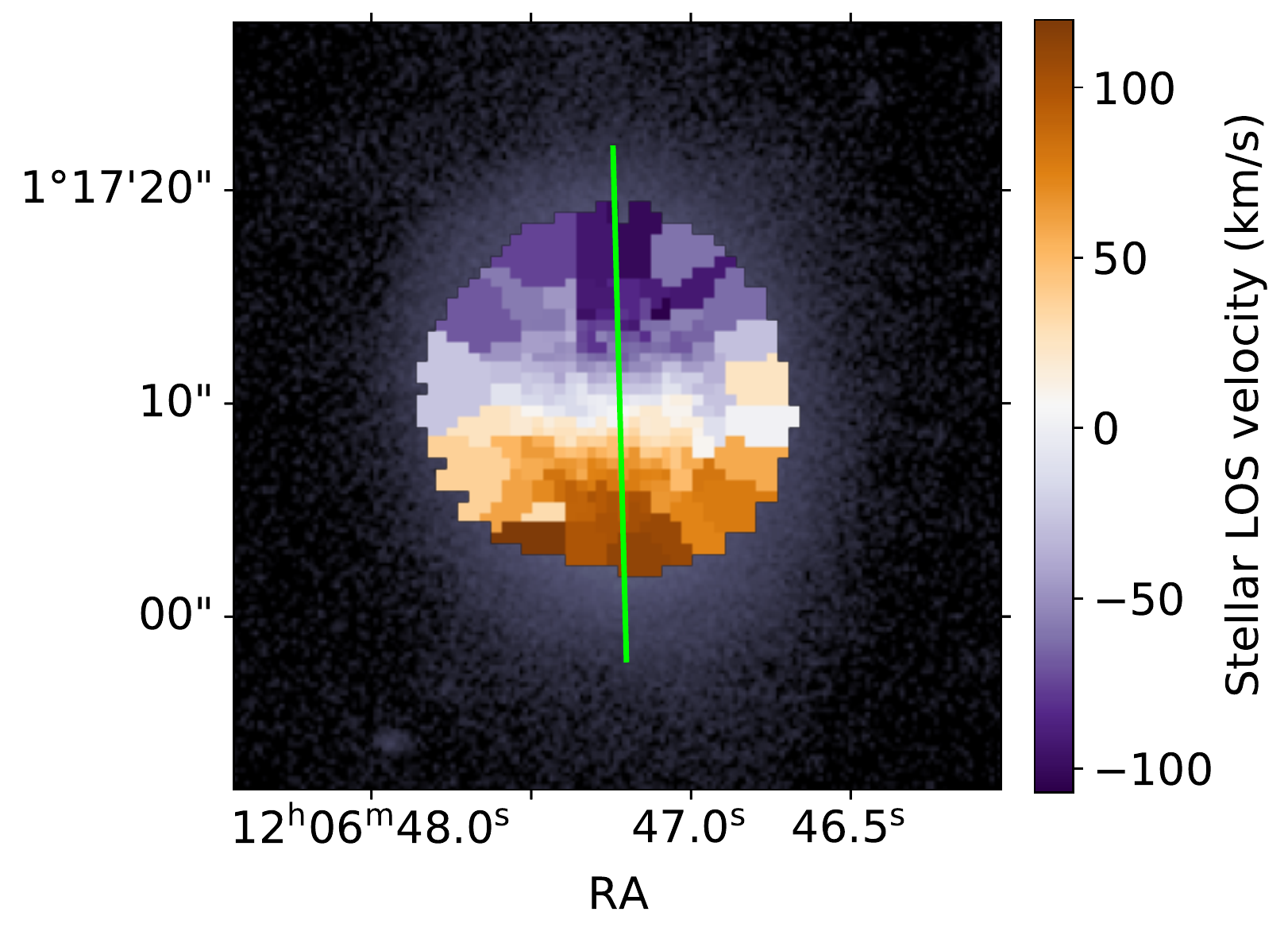}\hfill
   \includegraphics[width=.333\textwidth]{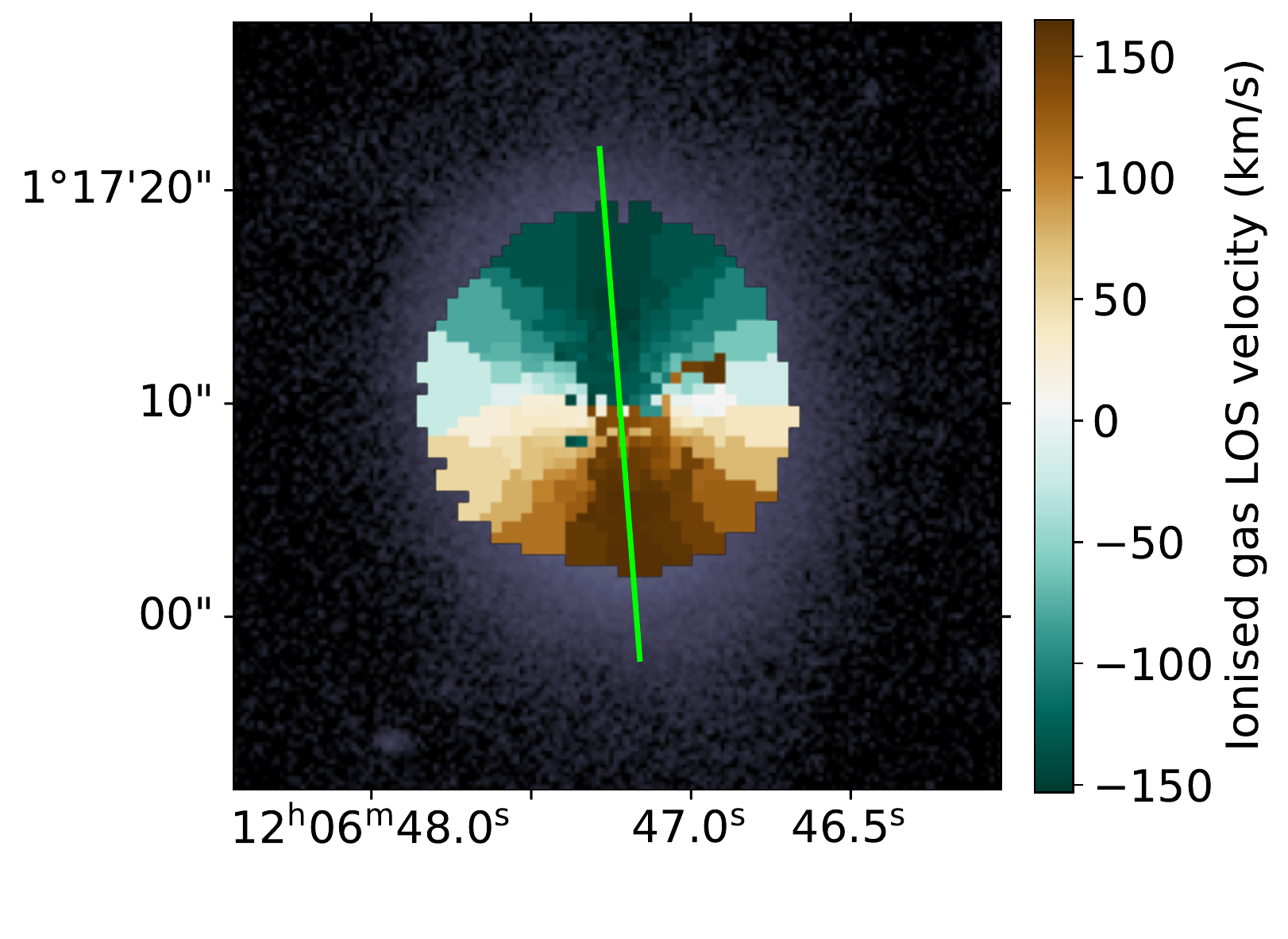}\hfill
   \includegraphics[width=.333\textwidth]{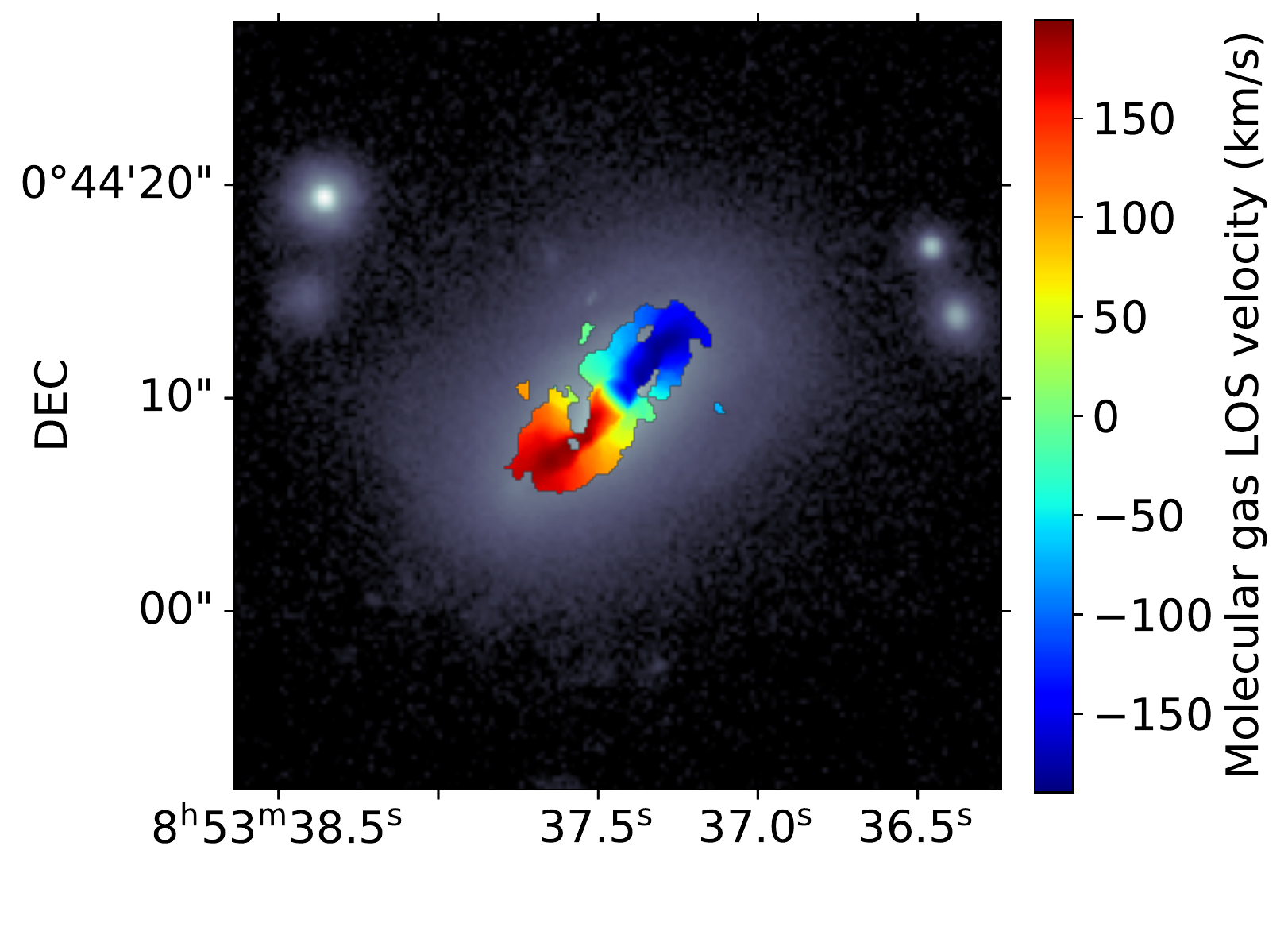}\hfill
   \includegraphics[width=.333\textwidth]{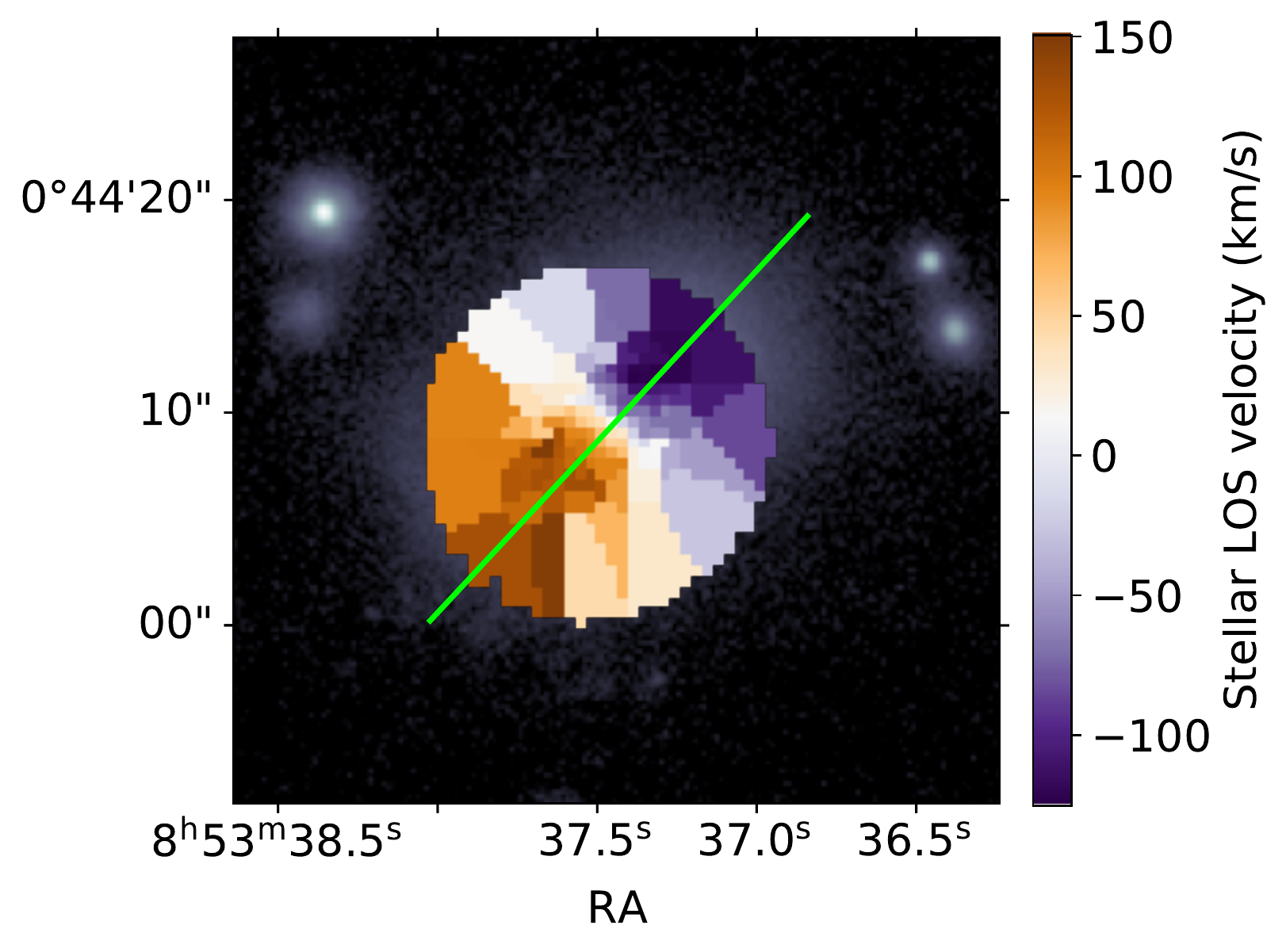}\hfill
   \includegraphics[width=.333\textwidth]{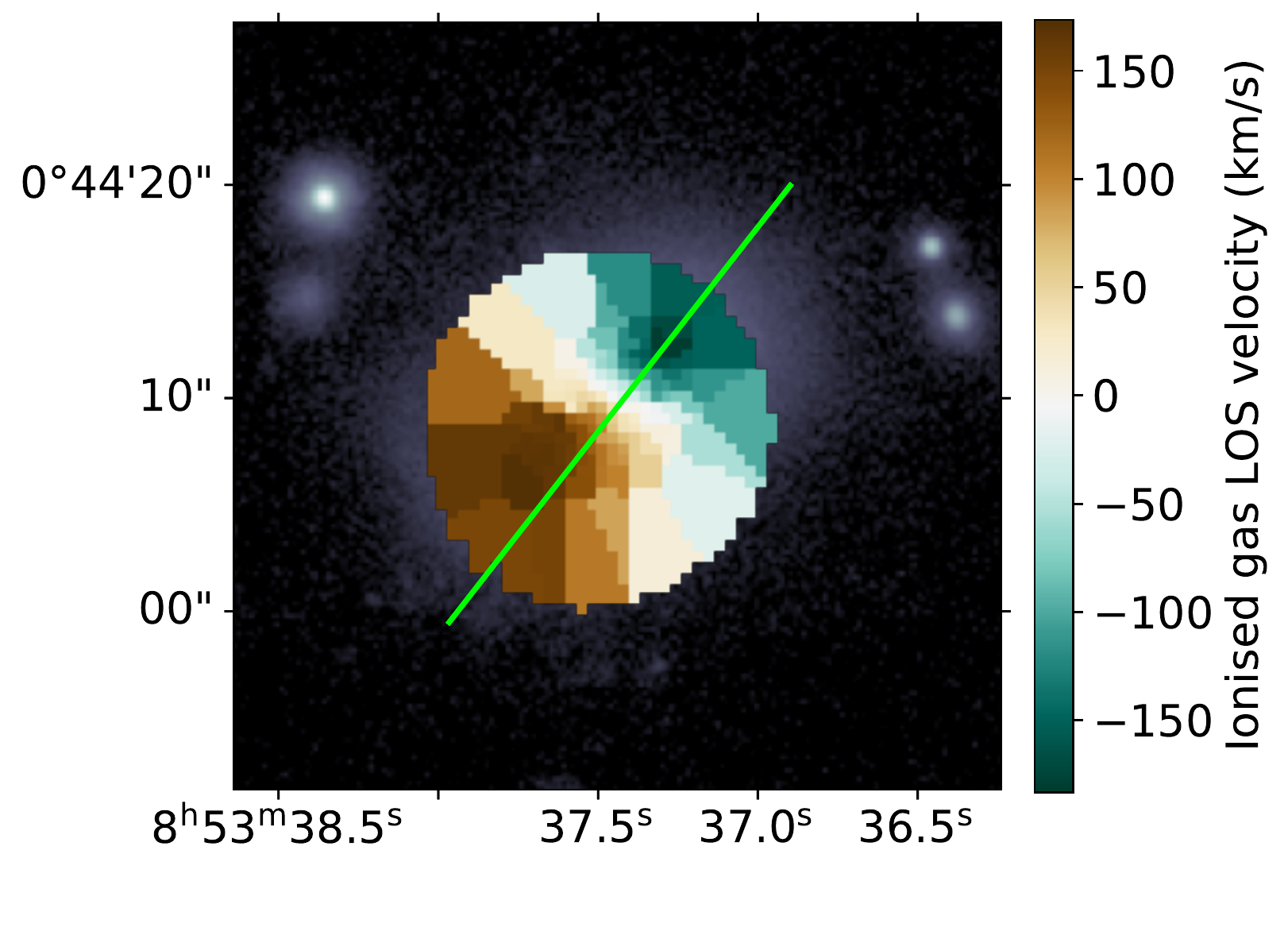}\hfill
   \caption{Kinematics of molecular gas (\textit{left}), stars (\textit{centre}) and ionised gas (\textit{right}), overlain on KiDS r-band log-normalised images. \textit{Top:} GAMA64646, \textit{middle:} GAMA272990, \textit{bottom:} GAMA622429.}
    \label{fig:SAMI_Kinematics_Plots}
\end{figure*}

Maps of molecular gas kinematics and fitted kinematic position angles were compared with kinematic maps of stars and ionised gas from SAMI DR3. Figure \ref{fig:SAMI_Kinematics_Plots} shows the first order moment maps for molecular gas from ALMA data and velocity maps for stars and ionised gas for the three ETGs subject to kinematic modelling, overlain on log-normalised r-band images of the ETGs. Table \ref{tab:Kinematic_PAs} shows the tabulated kinematic PAs from SAMI compared to the derived PAs for molecular gas from kinematic modelling (Section \ref{sec:Kinematic_Modelling}). 

\begin{table}
	\centering
	\caption{Kinematic position angles for molecular gas (from model fitting to ALMA data), stars and ionised gas (from SAMI DR3) for GAMA64646, 272990 and 622429.}
	\label{tab:Kinematic_PAs}
	\begin{tabular}{lccc}
		\hline
		Galaxy & Molecular & Ionised & Stars  \\
                       & Gas PA & Gas PA & PA  \\
                       & ($\degree$) & ($\degree$) & ($\degree$)  \\
            \hline
            GAMA64646   & 62.0 $\pm$ 0.5 & 62.0 $\pm$ 0.2 & 63.0 $\pm$ 1.0 \\
            GAMA272990 &  3.5 $\pm$ 0.8 & 4.5 $\pm$ 0.2 & 1.5 $\pm$ 1.5 \\
            GAMA622429 &  135.8 $\pm$ 1.6 & 142.5 $\pm$ 0.2 & 137.0 $\pm$ 1.6 \\
	\end{tabular}
\end{table}

The first-order moment maps for molecular gas show generally smooth distributions, and do not themselves indicate interaction or disturbance. However, some evidence for interactions derived from molecular gas velocities is apparent in some diagnostic plots shown in Figure \ref{fig:Diagnostic_Plots}, as discussed in Section \ref{sec:Results_KinMS} . The kinematics of molecular gas, stellar and ionised gas (presumably from star formation) are well-aligned (within a few degrees) in all cases. There is a reduction in peak line-of-sight rotation velocity for stars compared to molecular and ionised gas for GAMA64646 and GAMA272990, likely due to asymmetric drift where stars lag behind the local standard of rest and are significantly more pressure supported than molecular gas \citep{Stromberg46}. The stellar kinematic map for GAMA64646 (Figure \ref{fig:SAMI_Kinematics_Plots}) may show a faint bifurcation in rotation velocity profile at both ends of the stellar disc major axis, visible as localised decreases in velocity at the extreme ends of the disc close to the green major axis line. This could arise from disturbance such as a tidal interaction such as those shown in optical images in \citet{vandevoort18}, leading to alterations in stellar orbital velocities at the outer edge of the galaxy, or it could simply be a noise effect. The alignments for GAMA272990 are possibly because molecular gas, star formation and ionised gas all lie within the embedded disc discussed previously.


\subsection{ALMA Detections for GAMA177186}
\label{sec:GAMA177186}


GAMA177186 is classified as an elliptical galaxy within GAMA DR3, with very little ongoing star formation (SFR 0.05$^{+0.02}_{-0.01}$ M$_{\odot}$ yr$^{-1}$ from MagPhysv06). With ALMA, only a small detection of millimetre continuum (dust) emission was found at the centre of the ETG itself, with no molecular gas detected (S19). No SAMI DR3 SFR or kinematic maps are available for this ETG. Because of its elliptical morphology, it is possible that GAMA177186 formed from the merger of two similar-sized galaxies \citep{Xilouris04}. Any cool molecular gas present is likely to have been consumed by star formation as a result of the merger, leaving a "red and dead" elliptical galaxy.

However, detection of millimetre line and continuum emission is apparent in maps from the ALMA observation, $\sim$4 arcsec from the ETG centre. GALFIT \citep{Peng10} was used to confirm that the ALMA-detected offset source is unresolved, with shape parameters consistent with that of the ALMA synthesised beam. This type of source confusion is a known issue when using Herschel observations to select targets for ALMA observations, because of the beam sizes of the Herschel observations ($\sim$7 - 35 arcsec for a 3.5m aperture).  Multiple objects detected within a single Herschel beam can lead to offset or multiple ALMA detections not associated with the intended target \citep{Karim13}.

\begin{figure}
   \includegraphics[width=\columnwidth]{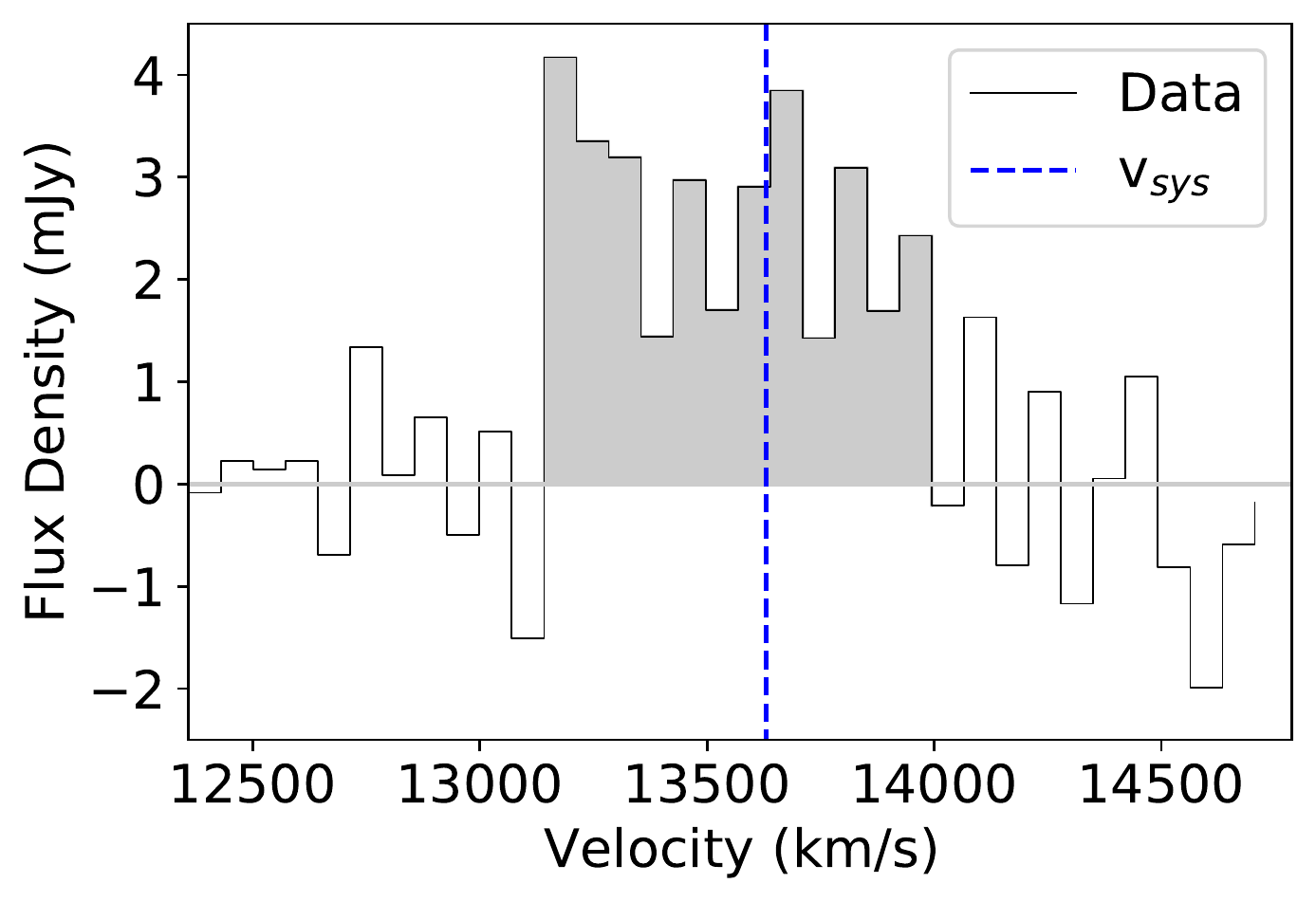}
   \caption{$^{12}$CO[2-1] spectrum for GAMA177186 offset object, with binning to 70.7 km/s. vsys is the recession velocity of GAMA177186 calculated from heliocentric optical redshift. v$_{sys}$ is the radio velocity of the ETG.}
    \label{fig:G177186_Spectrum}
\end{figure}

Figure \ref{fig:G177186_Spectrum} shows the ALMA-observed emission line spectrum for the offset object. The spectrum is binned by a factor of 7 to 70.7 km/s, and the emission line (shown in grey) is detected with flux/uncertainty of 7.6. The boundaries of the emission line were selected manually, on the basis of forming a coherent line with bin emission generally above the prevailing noise level. Integrating the spectrum yields a flux of 2.3 $\pm$ 0.3 Jy km/s, which is similar to the value reported in S19. The measured line width is approximately 840 $\pm$ 70 km/s (allowing an uncertainty of $\pm$1 bin), centred within 1 bin of the radio velocity of GAMA177186. The first order moment map for the object (S19) shows that the object is rotating, and the maximum circular velocity would then be approximately 420 $\pm$ 35 km/s. This is consistent with values for massive galaxies (e.g. stellar mass $>$10$^{11}$ M$_{\odot}$) \citep[e.g.][]{Kassin07}. An alternative explanation is that the wide emission line is from two aligned dusty satellite galaxies orbiting GAMA177186. The large line width alone is therefore not conclusive evidence of high redshift for the object, but this remains a possibility.

\begin{figure}
   \includegraphics[width=\columnwidth]{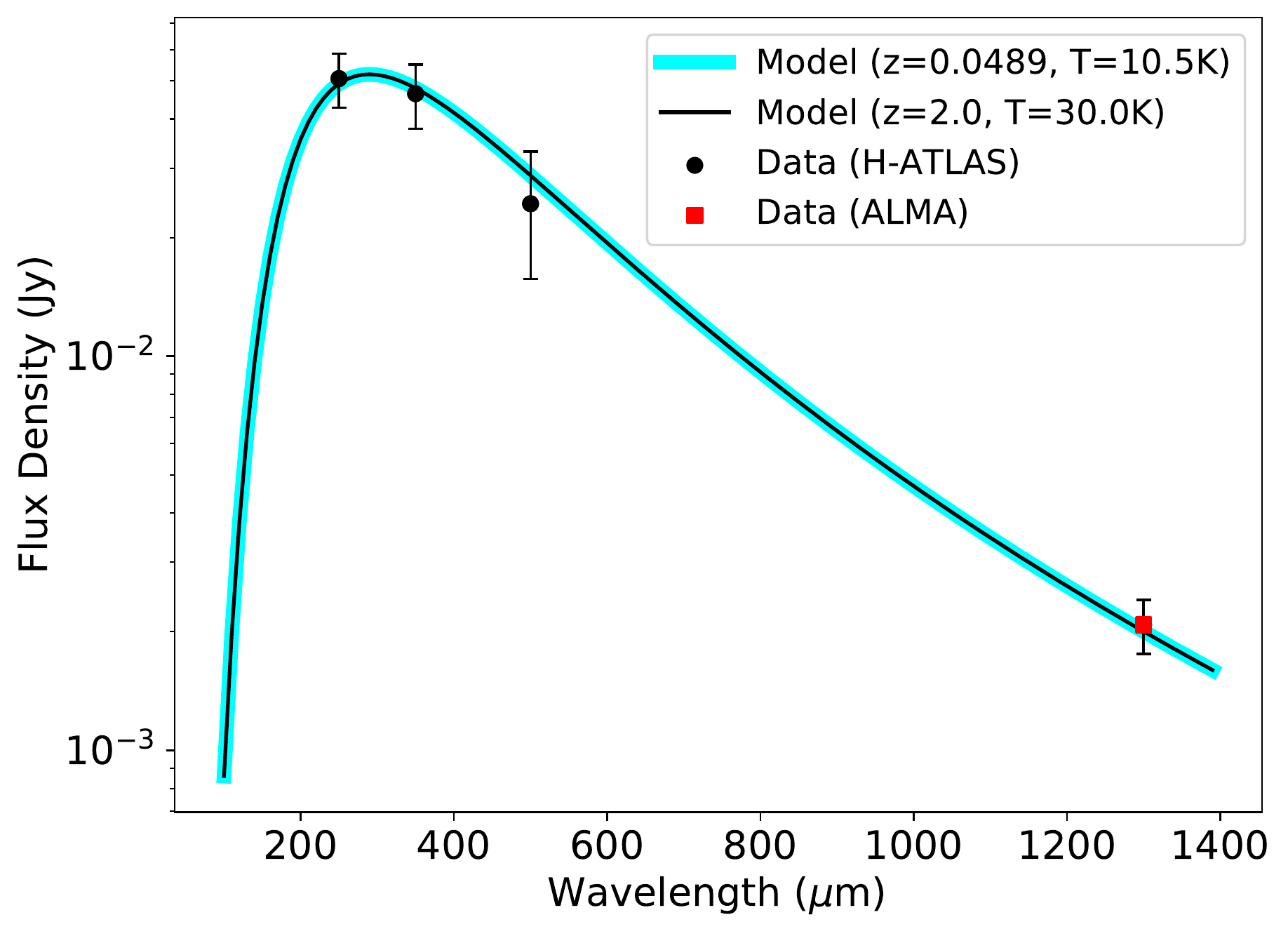}
   \caption{SED fitting of photometry from H-ATLAS and ALMA for GAMA177186 (assumed to be from the offset object.)}
   \label{fig:G177186_Dust_SED}
\end{figure}

Data from H-ATLAS DR1 \citep{Valiante16} shows sub-mm emission assigned to GAMA177186 detected in SPIRE passbands but not PACS passbands. When combined with photometry from the ALMA continuum map for GAMA177186 (S19), a consistent spectrum is obtained (Figure \ref{fig:G177186_Dust_SED}).  However, the derived flux density from the ALMA continuum map could be dominated by synchrotron emission from GAMA 177186 \citep{Condon92}, and the SPIRE measurements could be from dust within the galaxy that is undetected by ALMA due to MRS issues. The alignment of data points within the SED could therefore be coincidental. The analysis below is therefore for illustration only, and assumes that the Herschel-detected flux densities are from the offset object only.

\begin{figure}
   \includegraphics[width=\columnwidth]{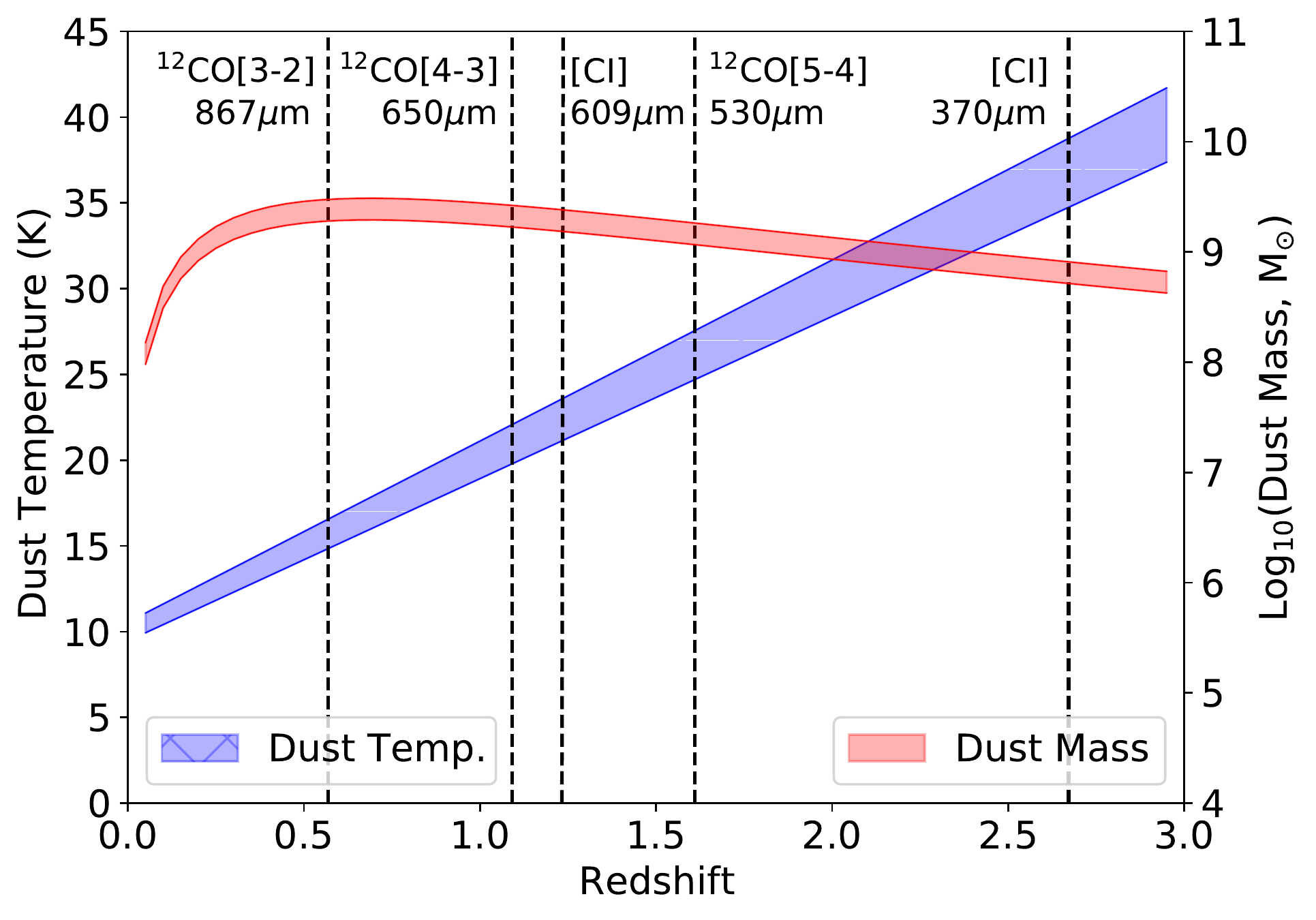}
	   \caption{Effect of assumed offset object redshift on fitted dust mass and temperature. Colour bands show 1$\sigma$ uncertainty range. Other emission lines possibly responsible for the detected line emisison are also shown, along with their rest wavelengths and associated redshift to produce the detected emission line.}
    \label{fig:G177186_Dust_Redshift}
\end{figure}

The spectrum in Figure \ref{fig:G177186_Dust_SED} was fitted with a single modified blackbody model representing a uniform dust population, with an emissivity coefficient ($\beta$) of 2, a mass absorption coefficient of 0.077 m$^2$/kg at 850 $\mu$m \citep[][and references therein]{Dunne11} and the ETG heliocentric redshift of 0.0476 from GAMA DR3. Colour corrections to measured flux densities at the fitted temperature were close to unity and therefore were not applied. This gave a dust mass of 10$^{8.1 \pm 0.5}$ M$_{\odot}$ and a temperature of 10.5 $\pm$ 0.6 K. Fitting the three SPIRE data points alone gave a dust temperature of 10.9 $\pm$ 1.8 K, so the ALMA-based flux density measurement is reasonably well aligned with those from SPIRE. Fitting was found to be equally successful for any positive redshift supplied to the model, with dust temperature increasing linearly with redshift and dust masses varying within an order of magnitude over the redshift range considered (Figure \ref{fig:G177186_Dust_Redshift}). The predicted dust temperature with the object at the redshift of GAMA177186 is consistent with dust temperatures found in the outskirts of M33 \citep{Thirlwall20} and other spiral galaxies where energy density of radiation fields is low (see also \citet{Popescu03}). However, this dust temperature is lower than the average generally found in ETGs ($\sim$25K)\citep{Smith12} and spiral galaxies ($\sim$20K) \citep{Davies12}. It is therefore possible that the ALMA-detected dust emission is from warmer dust at a higher redshift.

If the line emission shown in Figure \ref{fig:G177186_Spectrum} is from a source at high redshift, the emission is not from $^{12}$CO[2-1] but some other line whose emission is redshifted to produce the spectrum in Figure \ref{fig:G177186_Spectrum}. Possible candidates are shown in Figure \ref{fig:G177186_Dust_Redshift}, along with the required redshift to produce the detected emission line. [CII] emission (158$\mu$m) is also a candidate, requiring a redshift of 7.6 to produce the detected line. Although minimum gas temperatures for significant excitation to produce some of the lines discussed exceed the predicted dust temperatures at their associated redshifts, molecular gas and dust temperatures could be different so these lines could still be relevant.

\begin{figure}
   \includegraphics[width=\columnwidth]{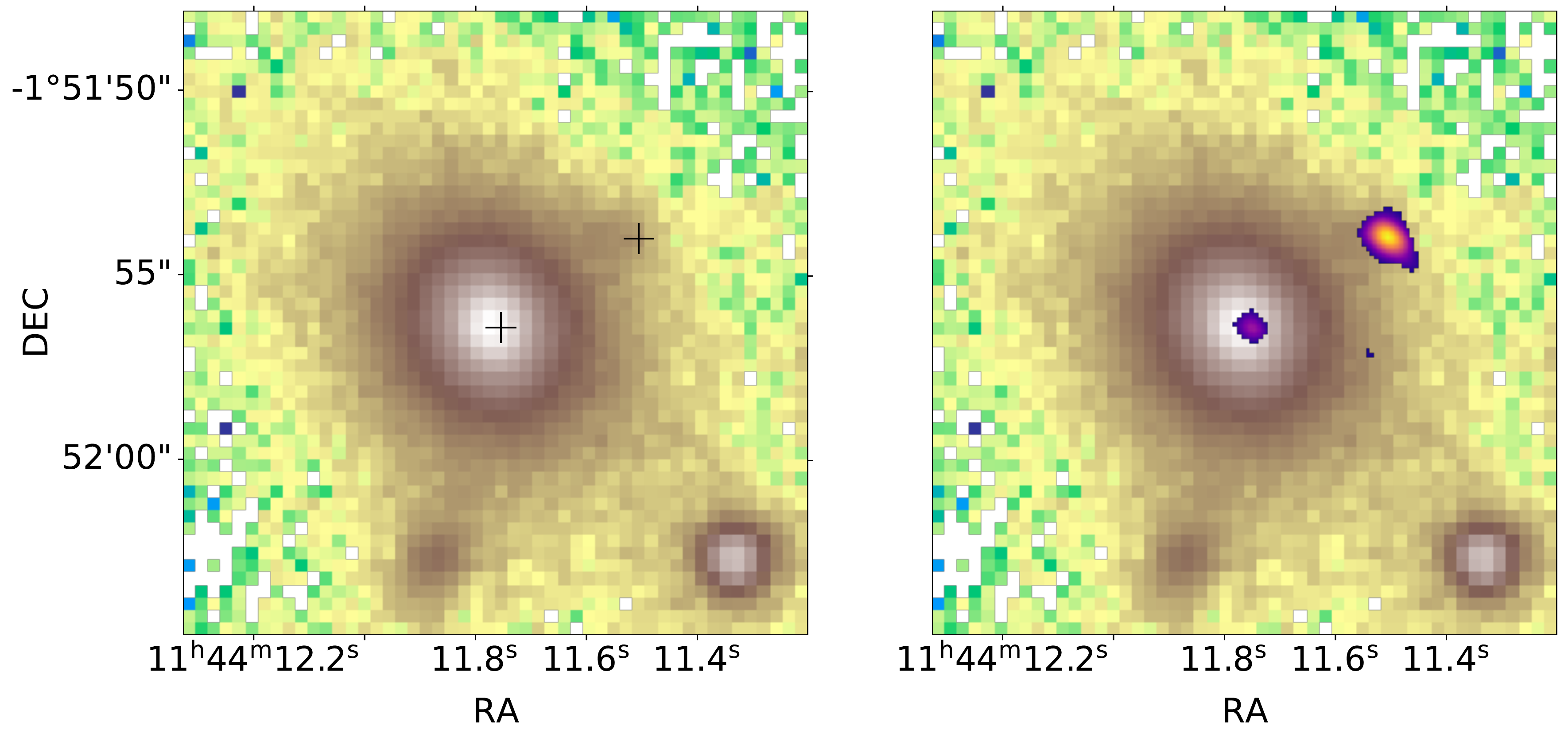}
   \caption{VISTA J, H and Ks-band co-added image (log-normalised) of GAMA177186 shown in yellow/brown/white. 
                \textit{Left}: centres of ALMA detections are shown as plus signs. \textit{Right}: Map of continuum detections are shown in blue/pink/yellow, overlain on the VISTA image.}
    \label{fig:G177186_J_H_K}
\end{figure}

On balance, the evidence for a high redshift for the offset object is not conclusive. One way of resolving this is to examine a very faint source apparent in VISTA \citep{Emerson04} J, H and Ks-band images from the VIKING survey \citep{Edge13} close to the ALMA-detected source. This source is not visible in VISTA Z or Y bands. When the pointing uncertainty of VISTA ($\sim$1 arcsec, \citet{Sutherland15}) is taken into account, the H/J/Ks-band emission and the ALMA detection are probably aligned (Figure \ref{fig:G177186_J_H_K}). Deeper infra-red images with mutiple passbands would confirm the presence of the feature photometrically and possibly alllow a photometric redshift to be estimated.

\subsection{ALMA Detections for GAMA622305}
\label{sec:GAMA622305}


Figure \ref{fig:WHAN} shows GAMA622305 as a retired galaxy with some emission lines present in its spectrum, at least some of which are attributed to ionising radiation from old stellar populations. Its SFR from GAMA DR3 (MagPhysv06) is 0.49$^{+0.02}_{-0.04}$ M$_{\odot}$ yr$^{-1}$. 

The ALMA observation of GAMA622305 (S19) also did not detect the dust content expected from Herschel observations, probably due to MRS issues discussed earlier (Section \ref{sec:ObsData}). However, a compact, elongated region of molecular gas was detected, with a central frequency close to that expected for $^{12}$CO[2-1] emission within the galaxy (Figure \ref{fig:r_gas_dust_overlay}). This source is offset from the galaxy centre, with a length of approximately 2 kpc. The first order moment map for this object (S19) shows a velocity distribution from low positive to low negative velocity along its length, with a velocity difference across the object of $\sim$100 km s$^{-1}$. The velocity profile is consistent with velocities of ionised gas shown in maps from SAMI, suggesting that the object is following the general rotation profile of the ETG.

\begin{figure}
   \includegraphics[width=\columnwidth]{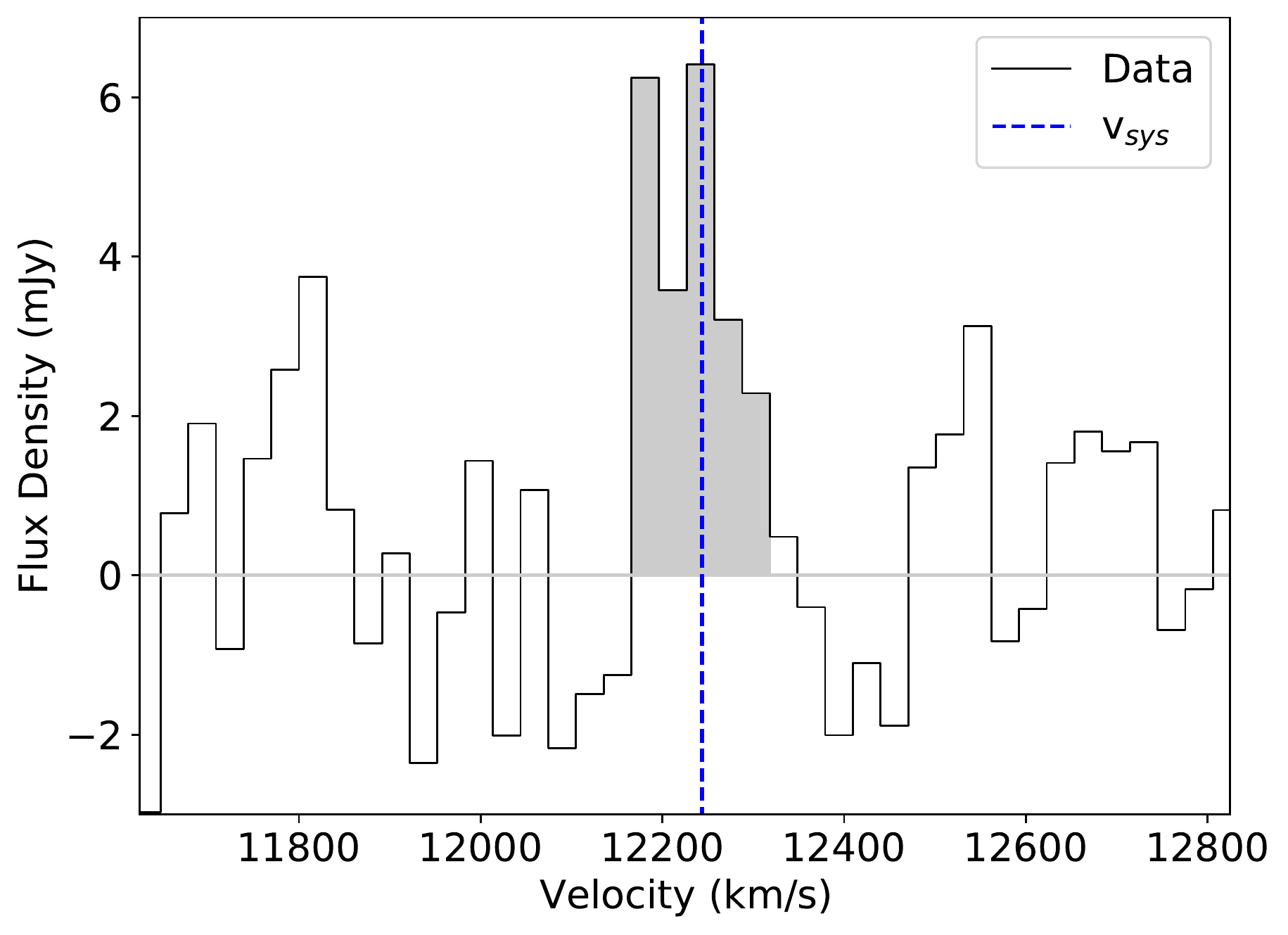}
   \caption{Spectrum for $^{12}$CO[2-1] detected in GAMA622305, with GAMA622305 velocity shown in relation to the emission line. v$_{sys}$ is the radio velocity of the ETG.}
    \label{fig:figE}
\end{figure}

The ALMA-observed spectrum for the object shown in Figure~\ref{fig:figE}  is binned by a factor of 3 (30.3 km s$^{-1}$ bins) compared to the spectrum shown in S19. Integrating the emission line (shaded grey) gives a flux density of 0.66 $\pm$ 0.14 Jy km s$^{-1}$, which is 0.1 Jy km s$^{-1}$ greater than that reported in S19 due to better definition of the emission line after binning. The emission line is detected with flux/uncertainty of 4.7.

\begin{figure}
   \includegraphics[width=\columnwidth]{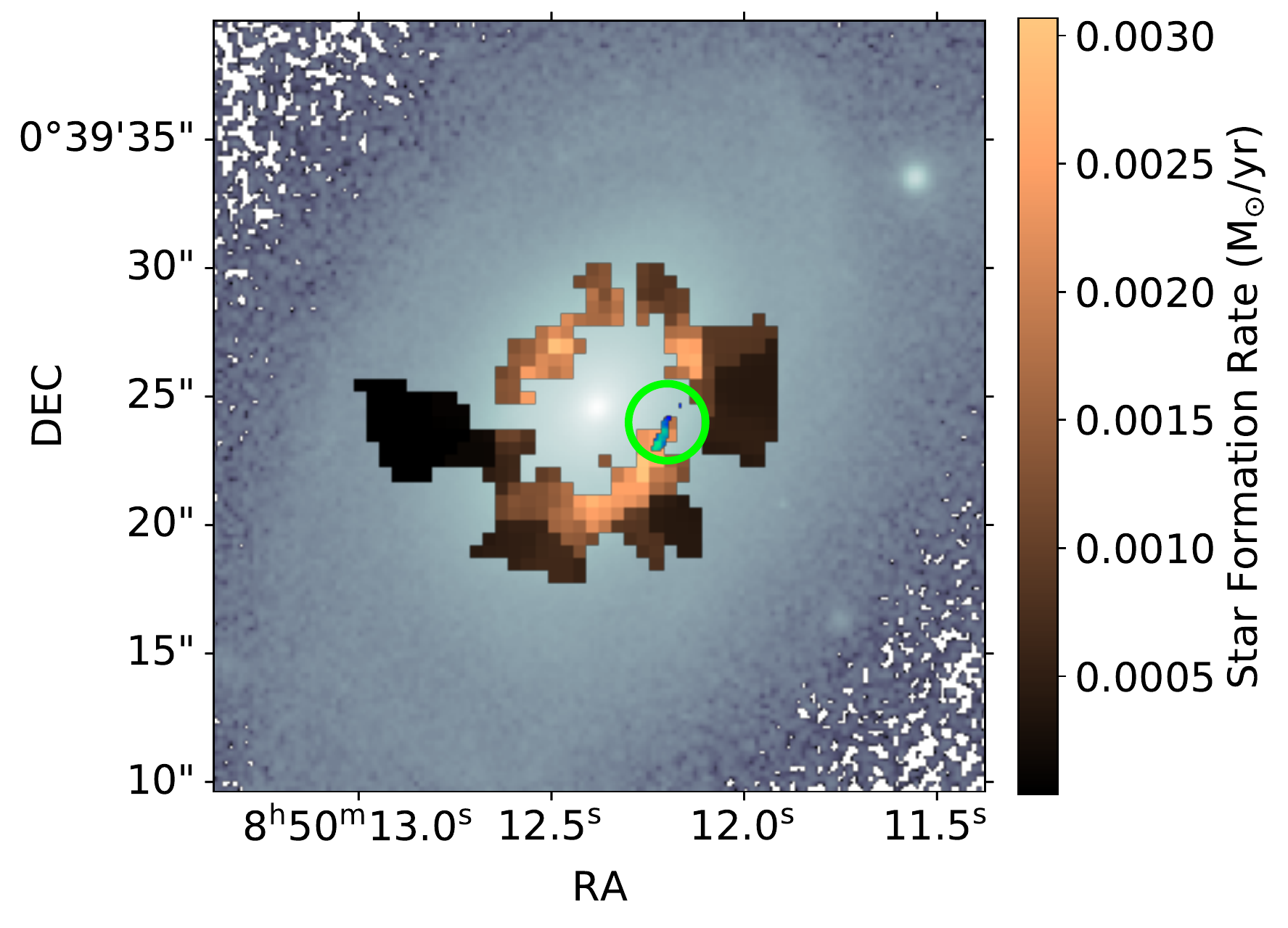}
   \caption{Star formation rate (SAMI, adaptive binning) and $^{12}$CO[2-1] emission (ALMA, shown and ringed in green) overlayed on KiDS r-band log-normalised image of GAMA622305.}
    \label{fig:figF}
\end{figure}

Figure~\ref{fig:figF} shows the location of the detected molecular gas in relation to a ring of SAMI DR3-detected star-forming regions within the galaxy. The spatial and velocity alignment of the molecular gas and ETG suggests that the molecular gas is associated with the star forming ring. The total SFR from this map is 0.28 M$_{\odot}$ yr$^{-1}$, which is less than the value from GAMA quoted earlier and suggests that either some star formation was not detected by SAMI or the MAGPHYS value is an over-estimate. The molecular gas mass of the object is estimated to be $\sim$8 $\times$ 10$^7$ M$_{\odot}$ by adjustment of the result in S19 to account for the increase in flux. This and the major axis dimension are both at least an order of magnitude greater than the most massive giant molecular clouds in other galaxies \citep[e.g.][]{Bolatto08}. The object could be a GMC complex \citep[e.g.][]{Kirk15} within the star forming ring, suggestive of clumpy star formation. 

\begin{figure*}

   \hspace{5em}\includegraphics[height=5.7cm]{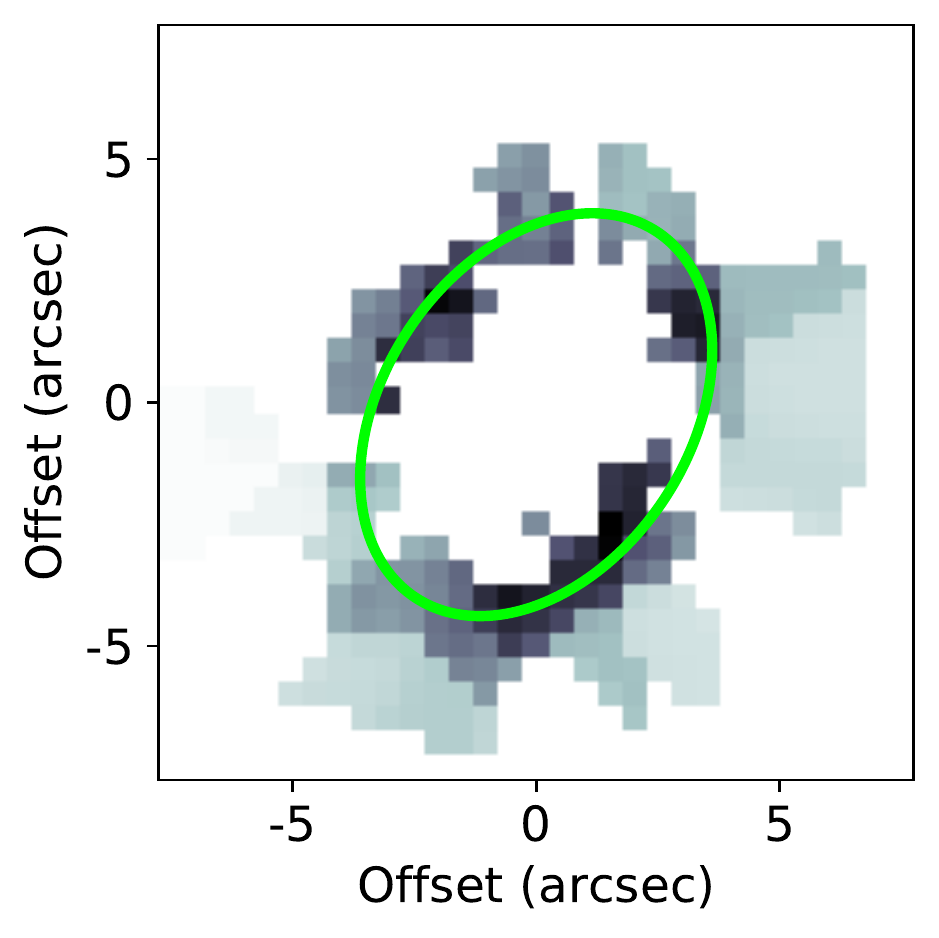}\hfill
   \includegraphics[height=5.7cm]{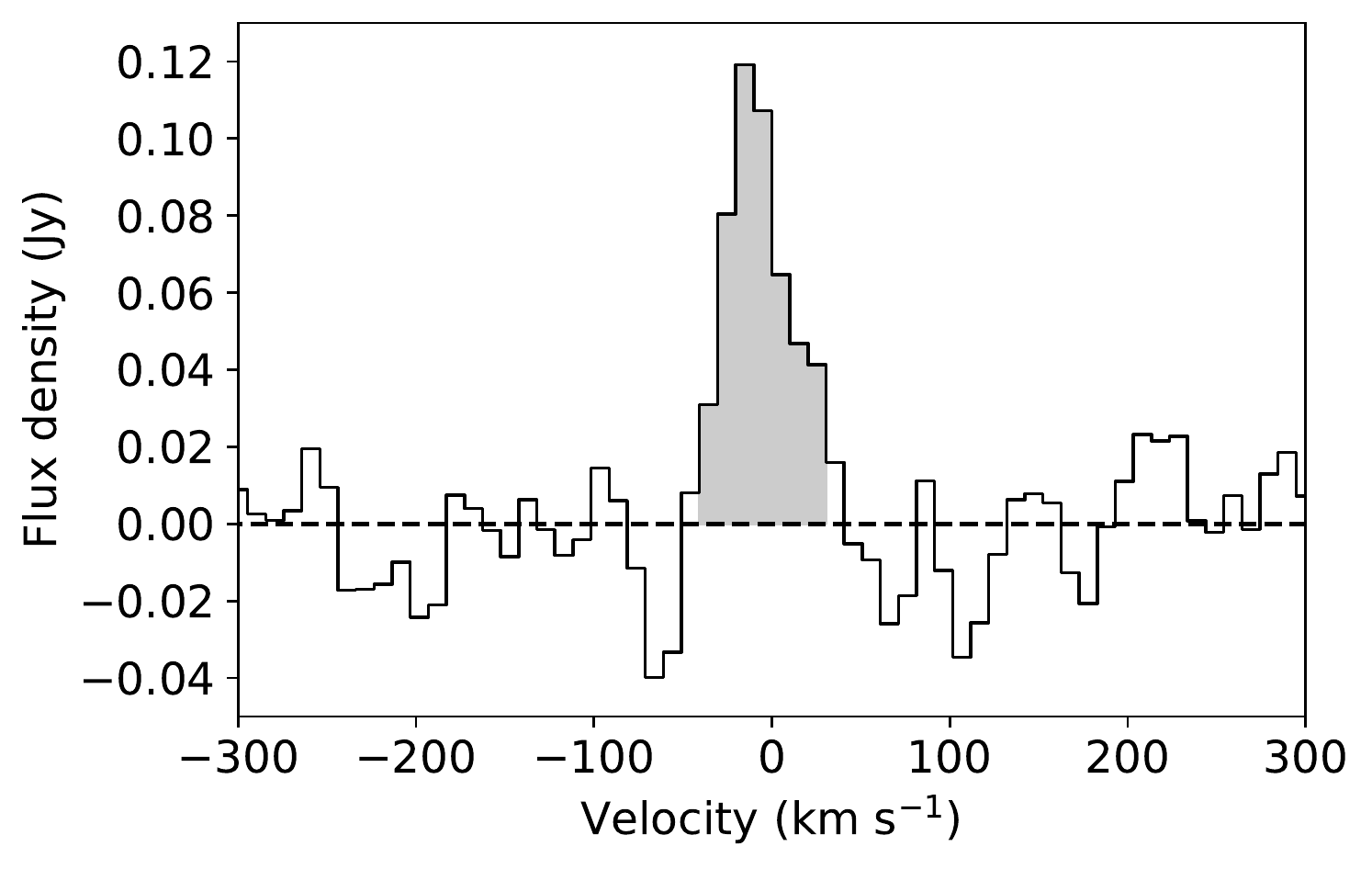}\hfill
   \includegraphics[height=5.7cm]{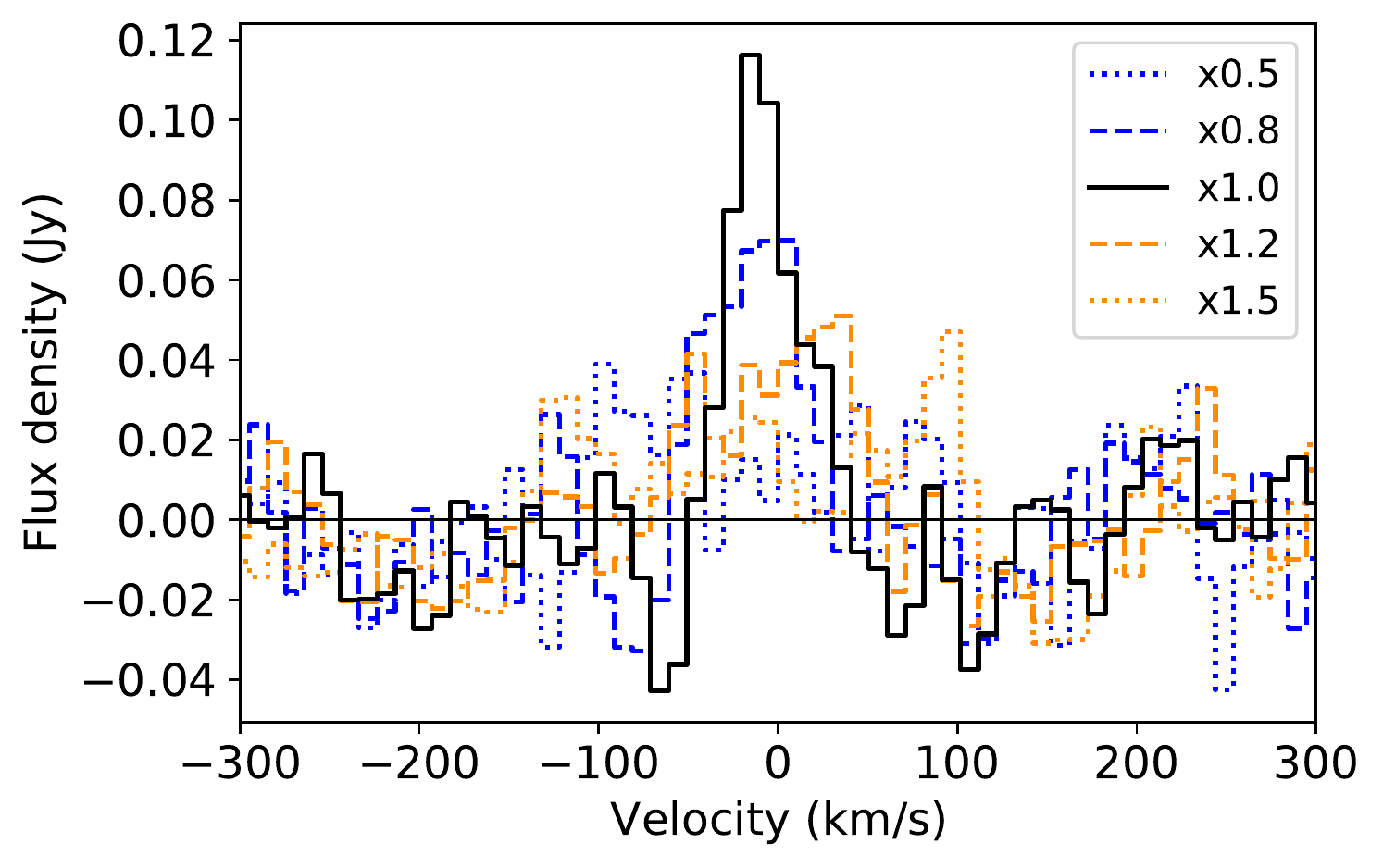}\hfill
   \includegraphics[height=5.7cm]{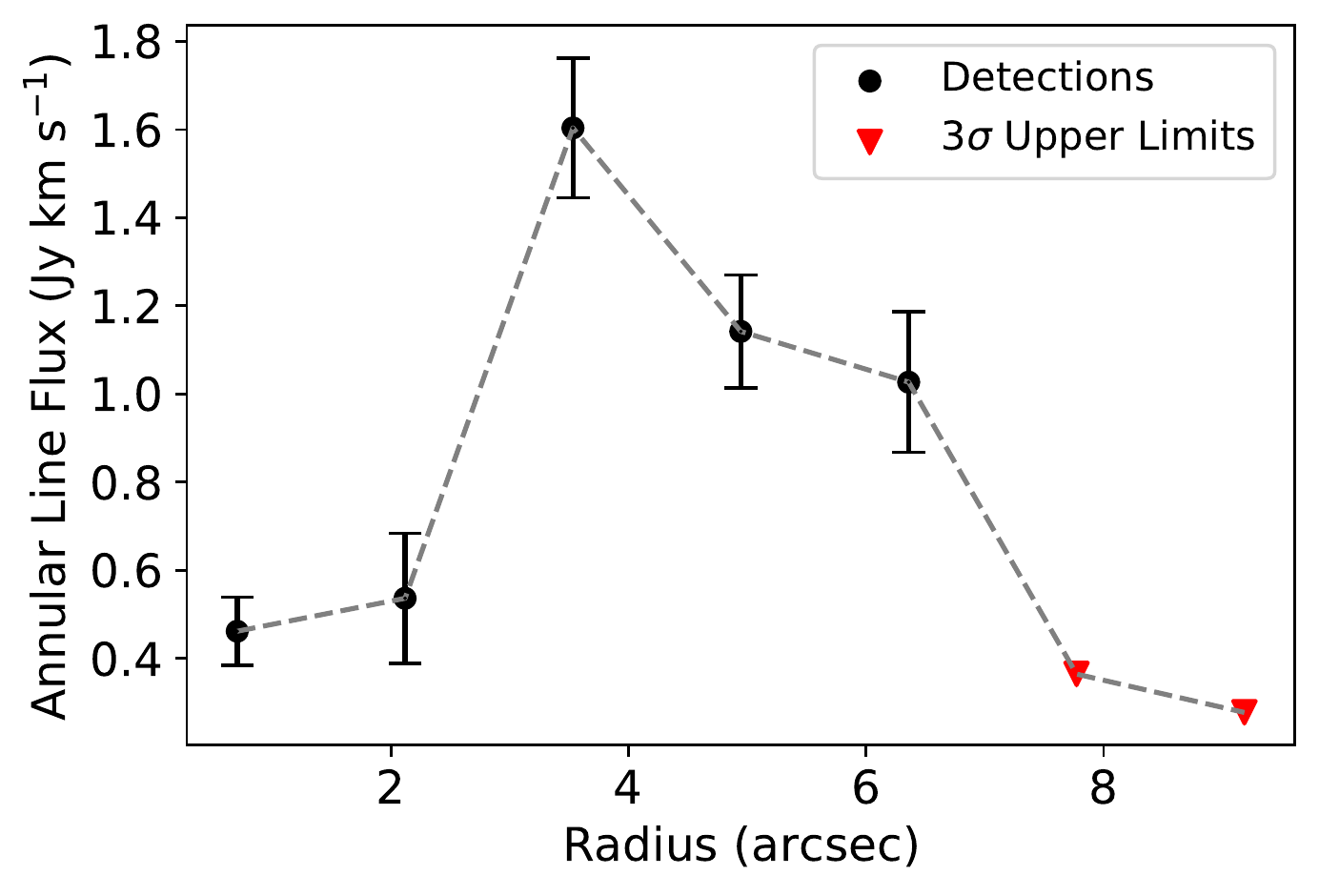}\hfill

   \caption{Analysis of GAMA622305 molecular gas distribution using \textsc{Stackarator}. \textit{Upper left}: star formation map (adaptive binning) from SAMI. Overlaid ellipse has a major radius of 4.5 arcsec, PA 146.5$\degree$, inclination 46.5$\degree$ from GAMA II r-band data. \textit{Upper right}: Emission line spectrum from stacking of whole data cube by Stackrator, using ionised gas velocity map from SAMI. Detected line is shown in grey. \textit{Lower left}: Spectra recovered when applying a multiple to the SAMI ionised gas velocity map. \textit{Lower right}: Fluxes in consecutive annular ellipses from Stackarator.}
    \label{fig:G622305_stackarator}
\end{figure*}

The presence of star formation in a ring within GAMA622305 suggests the presence of additional molecular gas to the feature discussed above, but this is not apparent during visual inspection of the data cube. However, emission in the data cube can be stacked using \textit{a priori} information on likely molecular gas velocity at defined spatial locations from an independent source, e.g. the ionised gas velocity map from SAMI. The \textsc{Stackarator}\footnote{https://github.com/TimothyADavis/stackarator} package (Davis et al., in prep.) does this for the cube as a whole to produce a single combined spectral line, and for concentric elliptical regions to obtain radial distributions of flux. Figure \ref{fig:G622305_stackarator} shows the results of using \textsc{Stackarator} on the whole data cube for GAMA622305. The combined spectral line (Figure \ref{fig:G622305_stackarator}, upper right) has an integrated flux of 4.98$\pm$0.34 Jy km s$^{-1}$, which is a factor of $\sim$7.5 greater than the flux of the molecular gas region shown in Figure \ref{fig:figF}. By extrapolation of the molecular gas mass quoted above for the highlighted region, the molecular gas mass in this ETG is then at least $\sim$6 $\times$ 10$^8$ M$_{\odot}$. 

To test whether this detection of additional flux using \textsc{Stackarator} is genuine, \textsc{Stackarator} was used to sample regions of the ALMA data cube coinciding spatially with ionised gas emission but at different velocities. This was achieved by scaling the SAMI velocity map by factors of 0.5, 0.8, 1.2 and 1.5. Figure \ref{fig:G622305_stackarator} (lower left) shows that the strength of the detected spectral line increases as the factor is increased towards 1, and then declines as the factor is increased further. Use of the unmodified SAMI velocity map for ionised gas with \textsc{Stackarator} therefore maximises the recovery of molecular gas emission, and the location of the recovered molecular gas emission coincides with ionised gas along the spatial and velocity axes.

Use of \textsc{Stackarator} to stack emission in elliptical annuli with a width of $\sim$2 ALMA beams (10 pixels, 1.2 arcsec) shows a sharp upturn in flux at $\sim$4 arcsec radius followed by a decline (Figure \ref{fig:G622305_stackarator}, lower right). The star forming ring has dimensions broadly consistent with this upturn (Figure \ref{fig:G622305_stackarator}, upper left), as expected if molecular gas and star formation are spatially coincident.

The optical image of GAMA622305 (Figure \ref{fig:r_gas_dust_overlay}) shows asymmetry, with the stellar distribution extended towards the lower left of the image. This ETG is therefore likely to have experienced a recent interaction that distorted it. New molecular gas could have been acquired during this event or existing molecular gas could have been disturbed, leading to the formation of a molecular gas ring with star formation as discussed previously. Formation of the ring by purely secular means \citep{Buta96} cannot be ruled out, but this ETG has been disturbed. Any pre-existing ring would have to be preserved through the disturbance.


\section{Discussion}

Of the five ETGs observed with ALMA, three (GAMA64646, 272990 and 622429) were found to contain massive ($>$10$^9$ M$_{\odot}$) molecular gas reservoirs. A fourth (GAMA622305) was found to contain a molecular gas feature consistent with a GMC complex and additional molecular gas probably in a ring structure, likely associated with a star-forming inner ring within the ETG. The main ALMA detection for the remaining ETG, GAMA177186, was from an object likely to be at a higher redshift than the target. The information presented for four of the observed ETGs with detected molecular gas can be used to infer the mechanisms responsible for their recent evolution. 

The information presented shows that all four ETGs have undergone disturbance of some kind. GAMA64646 appears to have undergone a tidal disturbance, probably leading to a star-forming ring in the SAMI observation and a tidal tail in optical images. GAMA272990 has an elliptical appearance in optical images which can be a product of a merger \citep[e.g.][]{Davis19}, with an asymmetric, star-forming embedded molecular gas disc. GAMA622429 has asymmetric molecular gas distribution and evidence of asymmetric disturbance in optical images, both coinciding with star formation detected in the SAMI observation. GAMA622305 appears to have undergone a tidal disturbance, which probably created a star-forming ring detected with SAMI that is coincident with the molecular gas feature detected by ALMA. It is possible that the molecular gas in these ETGs in low-density environments was acquired through merger activity at some point \citep{Davis11,Kaviraj12,Alatalo13}, but other interactions may have contributed to their observed appearance and behaviour.

The three massive molecular gas reservoirs detected by these ALMA observations all have a greater spatial extent than assumed for observation planning.  A median spatial extent for cool ISM of 1 - 2 kpc or half the optical (r-band) effective radius was assumed. Converting the radial scale factors and S\'ersic indices into effective radii containing 50\% of the $^{12}$CO[2-1] emission \citep{Macarthur03}, the effective radii of cool ISM for these three ETGs are in the range 2.8 - 5.2 arcsec (2.3 - 4.3 kpc), which is generally larger than the spatial extents originally assumed for molecular gas in ETGs. Future observation planning for cool ISM in dusty ETGs should be based on larger spatial extents, to ensure that all emission is accounted for. 

The five observed ETGs were selected mainly on the basis of having relatively large Herschel-observed dust masses and being free of strong AGN activity. Source confusion within the Herschel observation led to the selection of GAMA177186. GAMA622429 was selected in spite of having a strong AGN, probably due to incomplete line emission information to detect AGN activity. However, the ALMA observation of GAMA622429 has proved to be useful. In spite of this, the study of galaxies with strong AGN is not recommended for future studies of this type because estimates of SFR and other properties from optical data may not be reliable. ETGs with weak AGN activity (Figure \ref{fig:WHAN}) can be observed and studied successfully, because this study has shown that useful information can be gained from observing their widely-distributed ISM.

The gas depletion times and star formation rates for the two ETGs with massive molecular gas rotating discs (GAMA64646, 272990) can be compared with the Milky Way Galaxy (MWG), which is also acknowledged as being a GV galaxy. The MWG has less molecular gas \citep[6.5 $\times$ 10$^8$ M$_{\odot}$,][]{RomanDuval16}, similar star formation rates \citep[1.25$\pm$0.2 M$_{\odot}$ yr$^{-1}$,][]{Natale22}, and hence a shorter molecular gas depletion time (H$_2$ mass / SFR) of $\sim$0.5 Gyr. The MWG appears as a GV galaxy within colour-magnitude diagrams (e.g. \citet{Boardman20}, their Figure 1) along with many other dusty ETGs \citep[e.g.][]{Agius13}. These two ALMA-observed dusty ETGs with CO emission dominated by disc-like molecular gas reservoirs therefore differ from the MWG in terms of their future evolution. Their molecular gas reservoirs are not forming stars as fast as the MWG, which itself only has a modest SFR for its stellar mass. Reasons for this low star formation efficiency can be investigated in future studies.

The nature of the possible progenitors of GAMA272990 and GAMA622429 which may have undergone mergers cannot be studied with the data used here. Studies of the distributions of stellar populations and gas-phase metallicity using spectroscopy or IFU observations may assist in identifying the circumstances behind the mergers in more detail. For example, the stellar population ages for the embedded star-forming disc in GAMA272990 might be relatively young if the feature formed as a result of a recent merger, and gas-phase metallicities may be significantly different from stellar metallicities.

This work demonstrates that optical IFU observations can complement high-resolution interferometric observations (along spatial and frequency axes) of cool ISM in galaxies, but high-resolution observations provide unique information on the cool ISM itself and star formation arising from cool molecular gas. Star formation regions in rotating molecular gas discs from SAMI IFU observations align well with those determined from ALMA observations in this work. The Toomre stability criterion does seem to predict the places where molecular gas is observed to collapse to form stars, at least in this small sample of ETGs. However, because the SAMI star formation maps are designed to be clean but not complete as described above they can fail to identify star formation. Observations with high spatial and spectral resolution (e.g. interferometric) are also essential when investigating kinematic alignment of cool molecular gas, stars and ionised gas.

The evidence presented above indicates that the object offset from GAMA177186 could be a massive, dust- and gas-rich rotating object at relatively high redshift, but its redshift cannot be identified uniquely at present. It can be studied further if additional observations of sufficient depth become available at infra-red and other wavelengths.


\section{Conclusions}


This study has examined ALMA observations of $^{12}$CO[2-1] emission and continuum emission from five dusty (ISM-rich) visually-classified ETGs in low-density environments. These were selected from a complete sample within the GAMA equatorial fields with z $\leq$0.06. Axisymmetric or bisymmetric kinematic modelling was used to quantify structures and velocity profiles within the molecular gas, and differences between models and data were used to highlight possible events associated with the evolution of these ETGs. The aim was to investigate how resolved observations of ISM can assist in identifying the evolutionary mechanisms responsible for forming these ETGs. IFU maps and data from SAMI DR3 and data from GAMA DR3 provide further information. 

Four of the ETGs have massive ($\sim$few $\times$ 10$^8$ - few $\times$ 10$^9$ M$_{\odot}$), extended molecular gas reservoirs with radii effictive radii $\sim$ 3 - 5 kpc, apparent in the ALMA observations of molecular gas. All four of these ETGs appear to be undergoing evolution driven by disturbances. GAMA64646 and GAMA622305 appear to be evolving as a result of relatively mild external disturbances leading to environmental secular evolution. This is highlighted by faint distortions and tidal tails in optical images, and star-forming inner rings revealed by ALMA and IFU observations. GAMA272990 appears to be evolving due to a recent and more energetic interaction, leaving an elliptical appearance with an embedded star-forming molecular gas disc. GAMA622429 may have undergone a minor merger, due to asymmetry in its molecular gas distribution and asymmetric disturbance apparent in optical images. The remaining ETG (GAMA177186) was affected by source confusion, from an object which could be a massive gas- and dust-rich object at high redshift. Further observations are needed to establish the nature of this object conclusively. 

Overall, the observations and analysis indicate that a high proportion of dusty ETGs in low-density environments have undergone some kind of interaction as part of their recent evolution, although more observations of such targets are needed to verify and better quantify this result. Secular evolution over several gigayears can then complete the transformation of the ETGs from star-forming to passive. High-resolution interferometric observations are also shown to complement IFU observations, allowing a more complete picture of the evolution of these ETGs to be examined.


\section*{Acknowledgements}


The authors would like to thank the anonymous referee for their helpful comments on this work.

This work makes use of the following ALMA data: ADS/JAO.ALMA\#2015.1.00477.S. ALMA is a partnership of ESO (representing its member states), NSF (USA) and NINS (Japan), together with NRC (Canada), MOST and ASIAA (Taiwan), and KASI (Republic of Korea) in cooperation with the Republic of Chile. The Joint ALMA Observatory is operated by ESO, AUI/NRAO, and NAOJ. 

KiDS images are based on data products from observations made with ESO Telescopes at the La Silla Paranal Observatory under programme IDs 177.A-3016, 177.A-3017 and 177.A-3018, and on data products produced by Target/OmegaCEN, INAF-OACN, INAF-OAPD, and the KiDS production team, on behalf of the KiDS consortium. OmegaCEN and the KiDS production team acknowledge support by NOVA and NWO-M grants. Members of INAF-OAPD and INAF-OACN also acknowledge the support from the Department of Physics \& Astronomy of the University of Padova and of the Department of Physics of University of Federico II (Naples).

The Herschel-ATLAS is a project with Herschel, which is an ESA space observatory with science instruments provided by European-led Principal Investigator consortia and with important participation from NASA. The Herschel-ATLAS website is http://www.h-atlas.org.

GAMA  is  a  joint  European-Australasian  project  based around  a  spectroscopic    campaign    using    the    Anglo-Australian  Telescope.  The  GAMA  input  is  based  on  data taken  from  the  Sloan  Digital  Sky  Survey  and  the  UKIRT Infrared Deep Sky Survey. Complementary imaging of the GAMA regions  is  being  obtained  by  a  number  of  independent  survey  programs  including  GALEX  MIS,  VSTKiDS,  VISTA  VIKING,  WISE,  Herschel-ATLAS,  GMRTand  ASKAP  providing  UV  to  radio  coverage.  GAMA is funded by the STFC (UK), the ARC (Australia), the AAO,and  the  participating  institutions. The  GAMA  website  is http://www.gama-survey.org/.

The SAMI instrument was a collaborative development between the University of Sydney and the Australian Astronomical Observatory. The SAMI Galaxy Survey is supported by The ARC Centre of Excellence for All Sky Astrophysics in 3 Dimensions (ASTRO3D) and was previously supported by ARC Centre of Excellence for All Sky Astrophysics (CAASTRO).

This work has received funding from the European Union’s Horizon 2020 research and innovation programme under grant agreement No. 730562 (RadioNet).

\section*{Data Availability}

The data used within this work are or soon will be made available through ALMA, GAMA, KiDS, H-ATLAS and SAMI public data releases (https://almascience.nrao.edu/asax/, http://gama-survey.org/, http://kids.strw.leidenuniv.nl/, https://www.h-atlas.org/, https://sami-survey.org/). The specific data products generated for this paper will be shared on reasonable request to the lead author.



\bibliographystyle{mnras}
\bibliography{Paper2Refs} 




\appendix




\bsp	
\label{lastpage}
\end{document}